\documentclass[a4paper]{article}

\pdfoutput=1

\usepackage{a4wide}
\usepackage{authblk}
\usepackage{bbm}
\usepackage{amsmath}
\usepackage{amsfonts}
\usepackage{amssymb}
\usepackage{hyperref} 
\usepackage{apacite}
\usepackage{graphicx}
\usepackage{bbm}
\usepackage{epstopdf}
\usepackage{amsthm}
\usepackage[top=2in, bottom=1.5in, left=1in, right=1in]{geometry}
\usepackage{booktabs}
\usepackage{xspace}
\usepackage[english]{babel}
\usepackage{tikz}
\usepackage{color}
\usepackage{listings}
\usepackage{float}
\usepackage{geometry}
\usepackage{wrapfig}
\usepackage{enumerate}
\usepackage{watermark}
\usepackage[official]{eurosym}
\usepackage{subcaption}
\usepackage{todonotes}
\usepackage{authblk}

\graphicspath{{graphicspdf/}}

\title{Predicting the long-term citation impact of recent publications}
\date{\today}

\author[a]{Clara Stegehuis}
\author[b]{Nelly Litvak}
\author[c]{Ludo Waltman}
\affil[a]{Eindhoven University of Technology, Department of Mathematics and Computer Science, P.O. Box 513, 5600MB Eindhoven, The Netherlands}
\affil[b]{University of Twente, Department of Applied Mathematics, P.O.\ Box 217, 7500 AE Enschede, The Netherlands}
\affil[c]{Centre for Science and Technology Studies, Leiden University, P.O.\ Box 905, 2300 AX Leiden, The Netherlands}

\begin{document}
\maketitle

\section*{Abstract}
\addcontentsline{toc}{section}{Abstract}
A fundamental problem in citation analysis is the prediction of the long-term citation impact of recent publications. We propose a model to predict a probability distribution for the future number of citations of a publication. Two predictors are used: The impact factor of the journal in which a publication has appeared and the number of citations a publication has received one year after its appearance. The proposed model is based on quantile regression. We employ the model to predict the future number of citations of a large set of publications in the field of physics. Our analysis shows that both predictors (i.e., impact factor and early citations) contribute to the accurate prediction of long-term citation impact. We also analytically study the behavior of the quantile regression coefficients for high quantiles of the distribution of citations. This is done by linking the quantile regression approach to a quantile estimation technique from extreme value theory. Our work provides insight into the influence of the impact factor and early citations on the long-term citation impact of a publication, and it takes a step toward a methodology that can be used to assess research institutions based on their most recently published work.
\\\\
\textbf{Keywords}: citation analysis; citation impact; impact factor; prediction; quantile estimation; quantile regression.


\section{Introduction}\label{sec:question}

Citation counts are a popular indicator of the impact of scientific publications. In the evaluation of research institutions, bibliometric indicators based on the citations received by the publications of an institution often play an important role. However, the use of citation-based indicators is problematic when the impact of recent publications needs to be determined. One or two years after their appearance, most publications have received only a few citations. After one year, there are many publications with just one or two citations or even with no citations at all. Some of these publications may receive a lot of citations in later years, while others may attract hardly any attention in the future. This makes it difficult to determine the impact of recent publications. Nevertheless, research institutions often want their performance to be assessed based on their most recent work~\cite{bornmann2013problem}. In this paper, we therefore propose a model for making predictions of the impact that recent publications will have in the long term.

Our model predicts the long-term citation impact of a publication based on two variables, namely the impact factor of the journal in which the publication has appeared and the number of early citations the publication has received. Early citations are defined as citations received in the year in which a publication appeared or in the year thereafter. The two predictors that we use are easily available, and contrary to for instance the prediction approach proposed by~\citeA{wang2013quantifying}, they allow predictions to be made fairly soon after the appearance of a publication. Also, compared with other predictors that could be considered, such as the length of the reference list of a publication or the number of authors of a publication, the predictors that we use are relatively hard to manipulate. Earlier studies have shown that both impact factors and early citations are important predictors of future citations. In the next section, we will provide an overview of these earlier studies and we will discuss their relationship with our present work.

Earlier studies on citation impact prediction have often focused on providing a point estimate of the future number of citations of a publication. Given the high degree of uncertainty in citation impact predictions, we believe that it is more relevant to know the probability that a publication will receive a certain number of citations in the future. We therefore do not predict the average number of citations that a publication is expected to attract in the future, but instead we predict a probability distribution for the future number of citations of a publication. To predict this probability distribution conditional on a publication's impact factor and its early citations, we employ the technique of quantile regression introduced by~\citeA{koenker1978}.

We also study the relationship between our prediction model based on quantile regression and results from extreme value theory. To do so, we first use so-called Zenga plots, introduced recently by~\citeA{cirillo2013}, to establish that the citation distributions obtained in our analysis have a Pareto tail. This result then enables us to provide analytical insight into the behavior of the quantile regression coefficients for high quantiles. More specifically, we are able to link the regression coefficients to an estimator for the tail quantiles of a Pareto distribution developed in the framework of extreme value theory~\cite{dekkers1989}.

We use citation data for a large set of publications in the field of physics to test our prediction approach. The data is taken from the Web of Science database.

The paper is organized as follows. First, Section~\ref{sec:lit} discusses how our research relates to earlier work reported in the literature. Next, Section~\ref{sec:data} describes the data that were used in our analysis. Section~\ref{sec:model} then introduces our model for predicting the long-term citation impact of publications, conditional on impact factors and early citations. Section~\ref{sec:res} presents our empirical results. Sections~\ref{sec:coef} and~\ref{sec:k0} focus on the values obtained for the parameters of our model. Sections~\ref{sec:fit},~\ref{sec:comp},~\ref{sec:distribution}, and~\ref{sec:perf} address the fit of this model to the data and the predictive power of the model. Section~\ref{sec:Parest} studies the relationship between our model and results from extreme value theory. Section~\ref{ref:sensitivity} addresses the sensitivity of the parameters of our model to differences between fields of science, focusing on the fields of biology, chemistry and physics. Finally, Section~\ref{sec:conc} concludes the paper.


\section{Relation with earlier work}\label{sec:lit}

There is an extensive literature on modeling or predicting the number of citations of a publication based on all kinds of variables. An early study in this literature is the work by~\citeA{peters1994determinants}, who investigate the determinants of the citation impact of chemical engineering publications. More recent work in this literature is reported by, among others,~\citeA{walters2006predicting},~\citeA{haslam2008makes},~\citeA{fu2010using},~\citeA{wang2011mining},~\citeA{wang2012development},~\citeA{didegah2013determinants,didegah2013factors},~\citeA{bornmann2013percentile},~\citeA{yu2014citation}, and~\citeA{onodera_inpress_factors}. Various studies have also appeared in non-bibliometric journals~\cite<e.g.,>{haslam2010predicting,lokker2008prediction,mingers2010drivers}. Recent overviews of the literature on modeling or predicting citation impact are provided by~\citeA{didegah2013determinants,didegah2013factors} and~\citeA{onodera_inpress_factors}. Examples of variables that have been found to predict citation impact include the impact factor of the journal in which a publication has appeared, the type of study (e.g., original research vs.\ literature review), the number of pages of a publication, the number of references of a publication, the number of authors, institutions, and countries in a publication's address list, and the past performance of these authors, institutions, and countries.

It is important to emphasize that the objective of our work is different from the studies mentioned above. Like the above-mentioned studies, our interest is in predicting citation impact. However, our more specific interest is in using citation impact predictions in the evaluation of researchers, research groups, research institutions, and so on. In this specific context, many of the variables that have been found to correlate with citation impact should not be used for making citation impact predictions. Some variables have the problem that they can be easily manipulated. For instance, suppose researchers know that they will be evaluated based on the predicted citation impact of their publications, and suppose researchers also know that the citation impact of a publication will be predicted based on, for instance, the number of pages or the number of references of the publication. In that case, in order to be evaluated more favorably, it may be tempting for researchers to try to artificially increase the number of pages or the number of references of their publications. Hence, researchers may try to manipulate the variables that are used to make citation impact predictions. Other variables have the problem that they may lead to an undesirable self-reinforcing effect. For instance, suppose researchers are evaluated based on the predicted citation impact of their publications, and suppose the citation impact of a publication is predicted based on the citation impact of the earlier work of the authors of the publication. In that case, researchers who were successful in their older work will automatically be predicted to be successful also in their more recent work. This creates a self-reinforcing effect. Future success is determined by past success.

In order to avoid problems related to manipulation and self-reinforcing effects, we aim to predict the citation impact of a publication based on indicators that are available shortly after the publication's appearance and that can be considered to provide an impression of the value of the publication for the scientific community. Our focus is specifically on two indicators, namely the impact factor of the journal in which a publication has appeared and the number of citations a publication has received during the first year after its appearance. Other indicators that could be used are the number of downloads of a publication~\cite{brody2006earlier}, the number of readers according to a service such as Mendeley~\cite{thelwall_inpress_mendeley}, and other types of altmetric indicators~\cite{costas_inpress_altmetrics}. In this paper, however, our focus is on impact factors and early citations.

The use of early citations to predict long-term citation impact has been studied in various papers.~\citeA{glanzel1997possibility},~\citeA{burrell2003predicting},~\citeA{mingers2006modeling},~\cite{mingers2008exploring}, and~\citeA{wang2013quantifying} propose mathematical models that describe how publications accumulate citations over time. Using these models, they predict the citation impact of a publication in the longer term based on the publication's short-term citation history.~\citeA{adams2005early},~\citeA{levitt2011combined},~\citeA{bornmann2013percentile}, and~\citeA{wang2013citation} present empirical analyses of the correlation between short-term and long-term citation counts. Based on an analysis of publications from 1993 in six fields in the physical and life sciences,~\citeA{adams2005early} concludes that ``across reasonably large samples of research publications (not individual papers) it is possible to use initial citation counts predictively to index emerging quality relative to the field'' (p.\ 579).~\citeA{wang2013citation} performs an analysis of all publications from 1980 indexed in the Web of Science database and reports that the Spearman correlation between short-term citation counts and citation counts after 31 years ``rises from $0.266$ in year 1 to $0.756$ in year 3, and then slowly reaches 1 in year 31'' (p.\ 866). Studies based on correlations reveal general patterns. Individual publications may of course strongly deviate from these patterns. Extreme deviations can be observed in the case of so-called `sleeping beauties', which are publications that are hardly cited for a long time and then suddenly receive a lot of citations~\cite{vanraan2004sleeping}. The phenomenon of sleeping beauties illustrates the difficulty of making accurate predictions of long-term citation impact.

We are aware of three studies in which a comparison is made between the use of early citations and the use of impact factors for predicting longer-term citation impact.~\citeA{abramo2010citations} compare rankings of Italian universities based on citations and based on impact factors. They find that in certain fields, in particular in mathematics and in computer sciences, the ranking based on impact factors outperforms the ranking based on early citations in terms of the correlation with the ranking based on longer-term citation impact.~\citeA{levitt2011combined} propose a combined indicator of the impact of a publication that is obtained by taking a weighted average of the number of citations of a publication and the impact factor of the journal in which the publication has appeared. They report that in the case of a citation window of no more than one year the combined indicator provides a better prediction of the longer-term citation impact of publications in the field of economics than a straightforward indicator based only on citations. These results are in line with the findings of~\citeA{stern2014high} for publications in the fields of economics and political science.~\citeA{stern2014high} reports that shortly after the appearance of a publication the combined use of early citations and impact factors yields a better prediction of the longer-term citation impact of the publication than the use of early citations only.

We have now provided an overview of the literature that is most closely related to the research that we present in this paper. To make clear how our research contributes to the literature, let us summarize how our research differs from existing work:
\begin{itemize}
\item
Our interest is in predicting long-term citation impact based exclusively on impact factors and early citations. As mentioned above, we do not want to use variables that can be easily manipulated or that may cause self-reinforcing effects.
\item
Our interest is in predicting long-term citation impact within one or two years after the appearance of a publication. Unlike some earlier studies~\cite{levitt2008patterns,wang2013quantifying,wang2012development,wang2011mining}, we do not want to wait for five or more years before making predictions.
\item
Earlier work has shown that predicting long-term citation impact is a difficult task. Hence, it cannot be expected that the future number of citations of a publication can be predicted with a high degree of accuracy. Unlike most earlier work, our interest therefore is in predicting a probability distribution for the future number of citations of a publication. This probability distribution represents the uncertainty that we have about the number of citations a publication will receive in the future. We want this probability distribution to be predicted with a high degree of accuracy. We do not aim to provide a point estimate of the future number of citations of a publication.
\item
Most earlier studies \cite<an exception is>{glanzel1997possibility} actually consider a simplified version of the problem of predicting long-term citation impact. For instance, a study may consider the problem of predicting the number of citations that publications from 2005 have received by the end of 2014, where the prediction is based on information available at the end of 2006. However, how do we know whether the prediction model obtained for this problem will also work well when it is applied to a different time period? For instance, will the model also work well to predict, based on information available at the end of 2014, the number of citations that publications from 2013 will have received by the end of 2022? This essential question is left unanswered in most earlier studies, but it will be addressed in our research.
\end{itemize}


\section{Data}\label{sec:data}

In this paper, we use the in-house version of the Web of Science database of the Centre for Science and Technology Studies of Leiden University.
Only publications of the document types `article' and `review' are included in the analysis. For counting citations, author self-citations are excluded.

In order to estimate the coefficients of the regression model presented in Section~\ref{sec:model}, we use a specific set of publications. Because citation behavior differs between fields of science, only publications in the field of physics that were published in 1984 are included. In order to be included, a publication must belong to at least one of the following Web of Science subject categories: \textit{Applied Physics}, \textit{Fluids and Plasma Physics}, \textit{Atomic, Molecular and Chemical Physics}, \textit{Multidisciplinary Physics}, \textit{Condensed Matter Physics}, \textit{Nuclear Physics}, \textit{Particles and Fields Physics}, and \textit{Mathematical Physics}. Our entire data set includes 56,207 publications.

As already explained, we build a regression model with two predictors: The number of early citations of a publication and the impact factor of the journal in which a publication was published. In the rest of this paper, we will refer to these predictors as covariates. The number of early citations is defined as the number of citations received by a publication in the first year after its appearance. It is denoted by $c_1$. In our data set, $c_1$ is the number of citations that a publication has received before the end of 1985. Hence, for counting early citations, all citations received by a publication in 1984 and 1985 are included. The impact factor ($IF$) of a journal in 1984 is calculated as the average number of citations that publications published in the journal in 1982 and 1983 received in 1984. In the calculation of the impact factor of a journal, only publications of the document types `article' and `review' are taken into account, both on the citing side and on the cited side.

In Section~\ref{sec:perf}, we also consider two sets of publications in the field of physics published respectively in 1990 and 2000. These sets of publications are used to evaluate the predictive performance of our regression model. In Section~\ref{ref:sensitivity}, we address the sensitivity of the regression coefficients to a specific field. To this end, we use a set of publications in the fields of biology and chemistry published in 1984. 

%


\section{Regression model for quantile prediction}\label{sec:model}

As pointed out in Sections~\ref{sec:question} and~\ref{sec:lit}, our interest is not in providing a point estimate of the future number of citations of a publication. Instead, our focus is on predicting a probability distribution for the future number of citations of a publication. More specifically, our aim is to predict the quantiles of this probability distribution.

Formally, the $p$-th quantile $q(p)$ of a random variable $Y$ with distribution function $F$ is given by
\begin{equation*}
q(p)=F^{-1}(p)=\inf\{y:F(y)\geq p\}.
\end{equation*}
Hence, saying that a publication scores at the $p$-th quantile means that the number of citations of the publication is greater than or equal to the number of citations of a proportion of $p$ of all publications.

Our goal is to predict quantiles for the distribution of the number of citations received by a publication starting from the second year after its publication date. For example, in our data set of publications published in 1984, we consider quantiles for the number of citations received by a publication between January 1986 and December 2013. We refer to this as the future number of citations of a publication or, alternatively, as the long-term citation impact of a publication. In this section, we propose a model that predicts the quantiles of the long-term citation distribution of a publication, conditioned on the impact factor and the number of citations in the first year.

\subsection{Models for quantiles}\label{sec:regr}

Like~\citeA{ke2013}, we assume that each publication has a fitness factor $\eta$. This fitness factor gives information about the competitiveness of a publication relative to other publications in obtaining citations. The fitness factor depends on different factors $\phi_s$ that contribute to the success of a publication. The fitness factor is assumed to be a product of each of these factors raised to some power $\delta_s$:
\begin{equation}\label{eq:fitness}
\eta \propto \prod_s\phi_s^{\delta_s}.
\end{equation}

We predict the quantiles of the distribution of future citations, conditioned on the fitness factor. A higher fitness factor means that a publication has a higher competitiveness to obtain citations. Therefore one expects that the quantiles of publications with a higher fitness factor are higher. We assume that the $p$-th quantile of publications with fitness factor $\eta$, denoted by $q(p|\eta)$, is proportional to $\eta$:
\begin{equation*}
q(p|\eta)=\tilde{C}_p\eta.
\end{equation*}
Here $\tilde{C}_p$ is a constant independent of $\eta$ for each quantile $p$. 

We consider three definitions of the fitness factor: A definition based only on the impact factor $IF$, a definition based only on the number of citations in the first year $c_1$, and a definition based on both $IF$ and $c_1$. For clarity of notation, the exponents $\delta_1$ and $\delta_2$ from Eq.~\eqref{eq:fitness} are relabeled as $\beta$ and $\gamma$. The three definitions of the fitness factor are summarized in Table~\ref{tab:diffmod}. The constant $k_0$ is needed to account for publications that have zero citations after one year. We will discuss our choice for $k_0$ in Section~\ref{sec:coef}. 

\begin{table}[htbp]
  \centering
    \begin{tabular}{lllll}
    \toprule
    Model &$\eta$   &Quantile prediction\\
    \midrule
    Only $IF$ &    $\eta \propto IF^{\beta}$   & $q(p|IF)=\tilde{C}_pIF^{\beta_p}$ \\
    Only $c_1$ &    $\eta\propto(c_1+k_0)^\gamma$   &  $q(p|c_1)=\tilde{C}_p\left(c_1+k_0\right)^{\gamma_p}$\\
    Full model &   $\eta\propto IF^{\beta}(c_1+k_0)^\gamma$    & $q(p|IF,c_1)=\tilde{C}_pIF^{\beta_p}(c_1+k_0)^{\gamma_p}$ \\
    \bottomrule
    \end{tabular}
  \caption{Three models studied in this paper.}
  \label{tab:diffmod}
\end{table}

\subsection{Quantile regression}\label{sec:qregression}

In the model described in Section~\ref{sec:regr}, the logarithm of the quantiles is linear in the logarithm of the covariates. For example, for the full model in Table~\ref{tab:diffmod} we obtain
\begin{equation}\label{eq:percregr}
\ln\left(q(p|IF,c_1)\right)=\gamma_p\ln(c_1+k_0)+\beta_p\ln(IF)+{C}_p,
\end{equation}
where ${C}_p=\ln({\tilde{C}_p})$. Because the logarithm is an increasing function, the logarithm of the $p$-th quantile is equal to the $p$-th quantile of the log-transformed citation counts. This means that we can take the logarithm of the number of citations and then fit Eq.~\eqref{eq:percregr} to the quantiles of those values.

Equation~\eqref{eq:percregr} is fitted using quantile regression introduced by~\citeA{koenker1978}. While in standard least squares regression the sum of squared errors is minimized, quantile regression minimizes a different function. It solves
\begin{equation}\label{eq:quant}
\min_{\xi}\sum_i\rho_p(y_i-x_i\xi). 
\end{equation}
Here $y_i$ is the logarithm of the future number of citations of publication $i$, $x_i$ is the vector of log-transformed covariates corresponding to publication $i$, $\xi^T=\begin{bmatrix}C_p &\beta_p& \gamma_p\end{bmatrix}$, and the function $\rho_p$ is defined as
\begin{equation*}
\rho_p(z)=zp-z\mathbbm{1}_{z<0}=\begin{cases}
zp & \text{if }z\geq 0\\
z(p-1)&\text{if }z<0.
\end{cases}
\end{equation*}

Equation~\eqref{eq:quant} minimizes the difference between $p$ and the fraction of negative residuals~\cite{koenker1978}.
Hence, when all values are different, the empirical quantiles for the future number of citations are fitted precisely. In our case, many publications have the same number of citations, and therefore there can be small differences between actual and fitted quantiles (for an illustration, see the blue dots in Figure~\ref{fig:2000data} below).


\section{Quantile regression results}\label{sec:res}

In this section, we apply the regression model from Section~\ref{sec:model} to the data described in Section~\ref{sec:data}.

\subsection{Model coefficients}\label{sec:coef}

\begin{figure}[ht]
\centering
\begin{subfigure}[b]{0.32\textwidth}
\includegraphics[width=\textwidth]{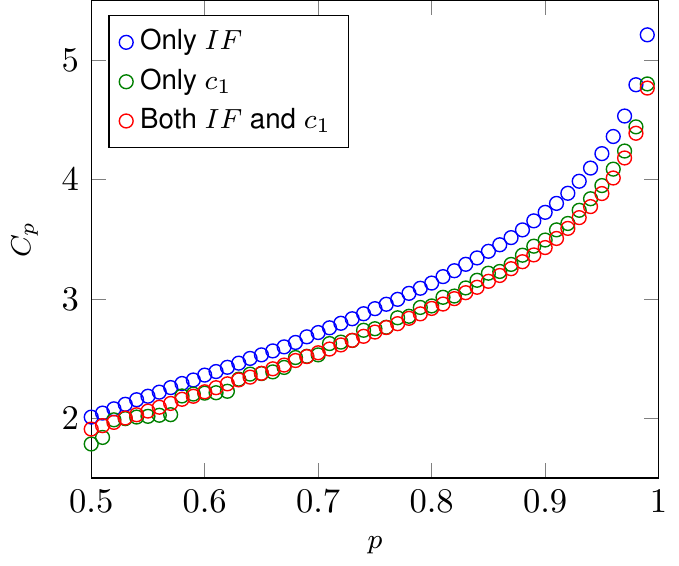}
\caption{$C_p$}
\label{fig:quc}
\end{subfigure}
\begin{subfigure}[b]{0.32\textwidth}
\includegraphics[width=\textwidth]{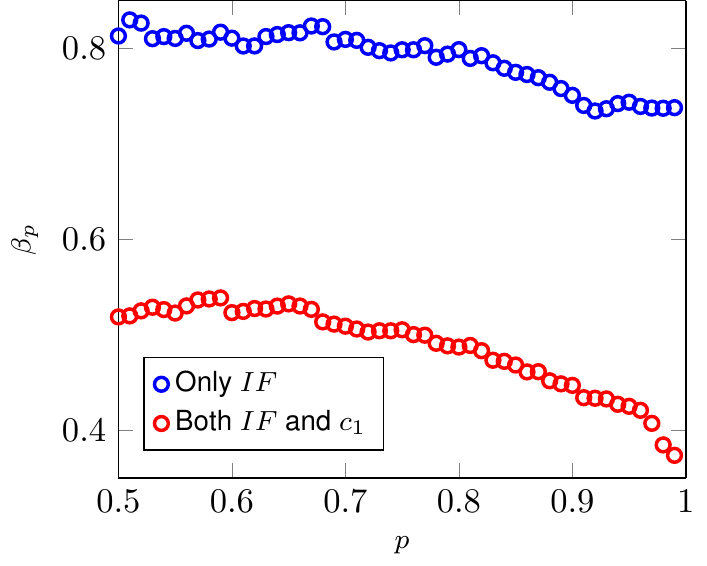}
\caption{$\beta_p$}
\label{fig:quab}
\end{subfigure}
\begin{subfigure}[b]{0.32\textwidth}
\includegraphics[width=\textwidth]{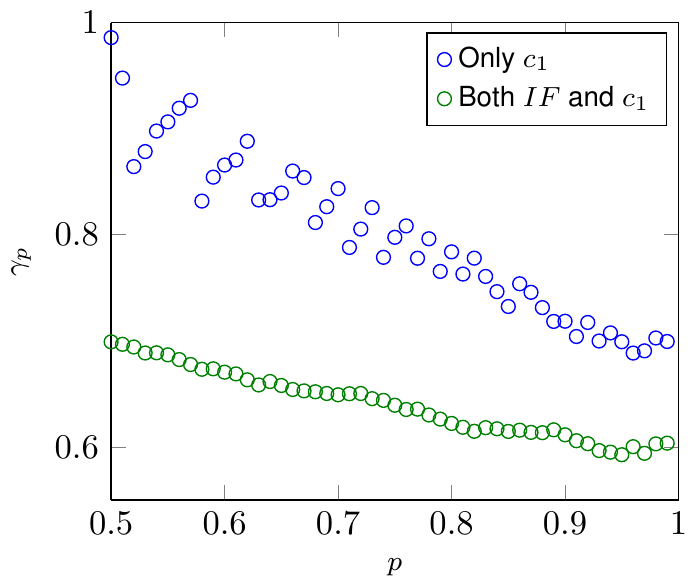}
\caption{$\gamma_p$}
\label{fig:qugam}
\end{subfigure}
\caption{Quantile regression coefficients for $p$-th quantile versus $p$ for the different models.}
\end{figure}

From a research evaluation perspective, high quantiles of citation distributions are especially important because interest often focuses on identifying high-impact research. For this reason, in our analysis we consider the $0.50$-th up to the $0.99$-th quantile. Figures~\ref{fig:quc},~\ref{fig:quab} and~\ref{fig:qugam} show the parameters $C_p, \beta_p$ and $\gamma_p$ resulting from the quantile regression. The coefficients are shown for the three different versions of the model listed in Table~\ref{tab:diffmod}. 

The coefficient $C_p$ in Figure~\ref{fig:quc} is increasing in $p$ for all three versions of the model. This is to be expected, because the quantiles are nondecreasing in $p$. We also see that $C_p$ is a convex function. For the higher quantiles, $C_p$ grows faster in $p$ than for the lower quantiles. This indicates that for example the $0.98$-th and the $0.99$-th quantile are further away from each other than, say, the $0.60$-th and the $0.61$-th quantile. 
In Section~\ref{sec:Parest}, we will explain the behavior of $C_p$ more precisely using quantile estimators for Pareto-tailed distributions.

The coefficients $\beta_p$ and $\gamma_p$ are decreasing in $p$. Hence, the impact factor and the number of citations in the first year have less influence on the long-term citation impact of highly cited publications than on the long-term citation impact of publications with an average number of citations.

\subsection{Influence of $k_0$}\label{sec:k0}

\begin{figure}[htb]
\centering
\begin{subfigure}{0.32\textwidth}
\includegraphics[width=\textwidth]{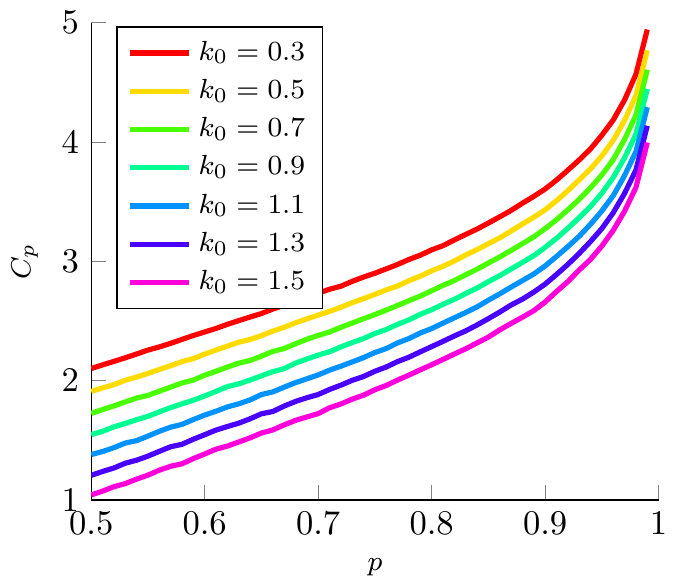}
\caption{$C_p$}
\label{fig:infk0}
\end{subfigure}
\begin{subfigure}{0.32\textwidth}
\includegraphics[width=\textwidth]{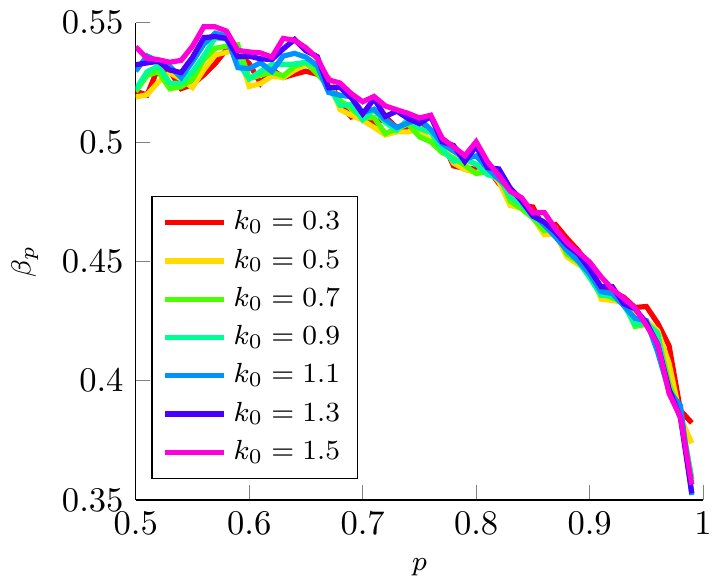}
\caption{$\beta_p$}
\label{fig:infk0ab}
\end{subfigure}
\begin{subfigure}{0.32\textwidth}
\includegraphics[width=\textwidth]{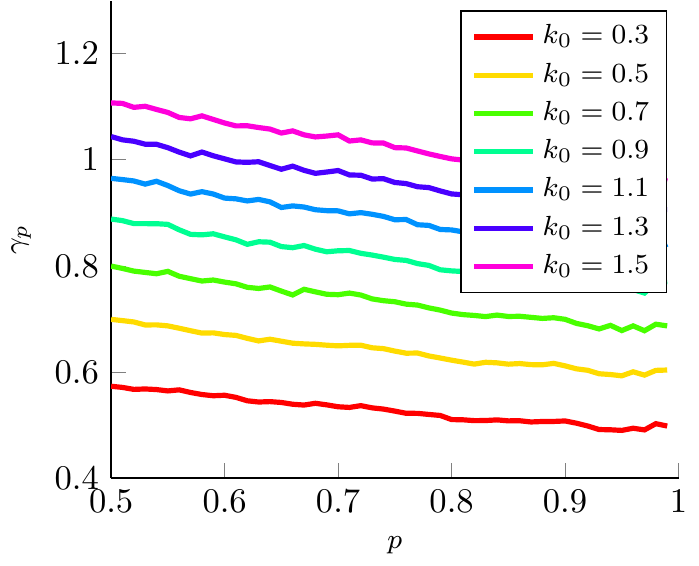}
\caption{$\gamma_p$}
\label{fig:infk0gam}
\end{subfigure}
\caption{Quantile regression coefficients for $p$-th quantile versus $p$ for $k_0$ ranging from $0.3$ to $1.5$.}
\end{figure}

The parameter $k_0$ in the models listed in Table~\ref{tab:diffmod} is not fitted in the quantile regression. To get an understanding of the influence of $k_0$ on the regression coefficients, quantile regression is used to obtain the coefficients $C_p$, $\beta_p$, and $\gamma_p$ for several values of $k_0$. Figures~\ref{fig:infk0},~\ref{fig:infk0ab}, and~\ref{fig:infk0gam} show the values of the regression coefficients for $k_0=0.3$ to $1.5$ for the full model. We see that $k_0$ has hardly any influence on $\beta_p$. Also, $k_0$ does not have much influence on $\gamma_p$ and $C_p$. Essentially, these coefficients increase or decrease by a constant value if $k_0$ changes. We use the value of $k_0$ that minimizes the sum of the squared difference between the fraction of publications with fewer citations than the predicted $p$-th quantile and $p$. This results in $k_0=0.5$.

\subsection{Fit of the models}\label{sec:fit}

\begin{figure}[htbp]
\centering
\begin{minipage}{0.4\linewidth}
\includegraphics[width=\textwidth]{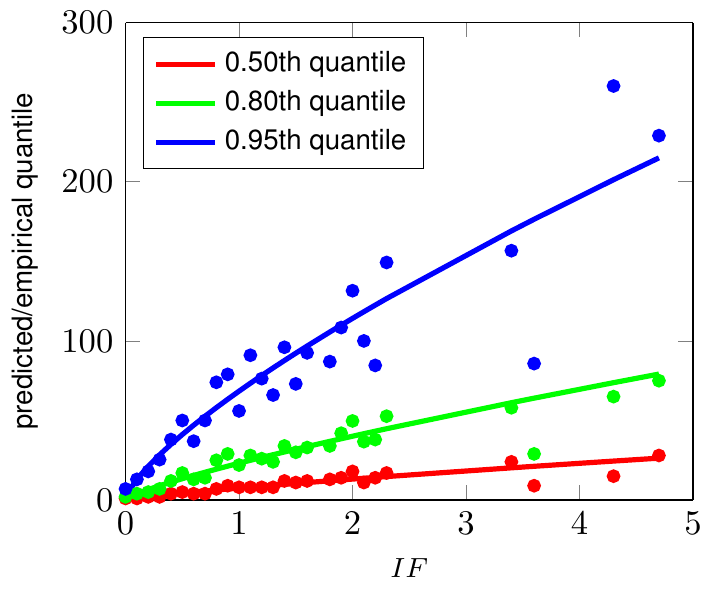}
\caption{Predicted value (solid line) and empirical value (dots) of the $0.50$-th, $0.80$-th, and $0.95$-th quantile versus $IF$ for the model that uses only $IF$.}
\label{fig:onlyIFfit}
\end{minipage}
\hspace{0.1cm}
\begin{minipage}{0.4\linewidth}
\includegraphics[width=\textwidth]{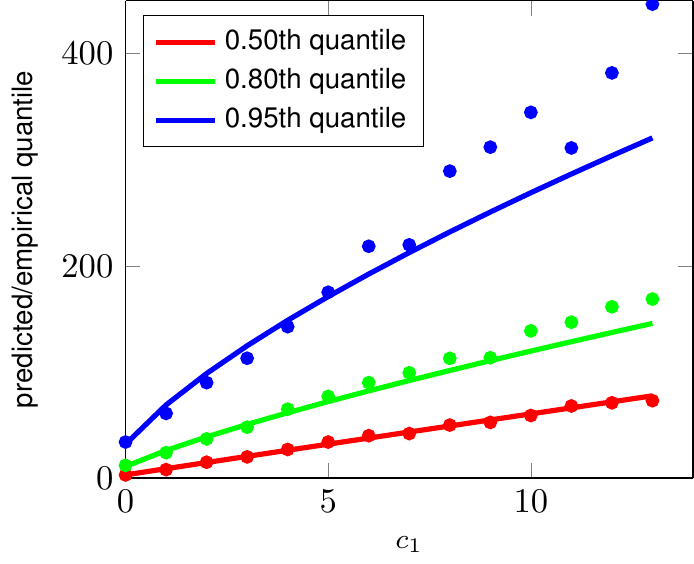}
\caption{Predicted value (solid line) and empirical value (dots) of the $0.50$-th, $0.80$-th, and $0.95$-th quantile versus $c_1$ for the model that uses only $c_1$.}
\label{fig:onlyc1fit}
\end{minipage}
\end{figure}

We now investigate the fit of the models listed in Table~\ref{tab:diffmod} to the data. The fit of the model that uses only the impact factor is illustrated in Figure~\ref{fig:onlyIFfit}. For every value of the impact factor, the predicted $0.50$-th, $0.80$-th, and $0.95$-th quantiles are shown as solid lines and the empirical quantiles are shown as dots. It is clear that the predicted and empirical quantiles may differ a lot. In a similar manner, the fit of the model that uses only the number of early citations is illustrated in Figure~\ref{fig:onlyc1fit}. For the $0.50$-th quantile, the empirical and the predicted quantiles almost overlap. For the $0.80$-th and the $0.95$-th quantile, the model fits well for publications with a small number of citations in the first year, but it underestimates the quantiles for publications with a large number of early citations.

\begin{figure}[htb]
\centering
\begin{minipage}{0.4\linewidth}
\includegraphics[width=\textwidth]{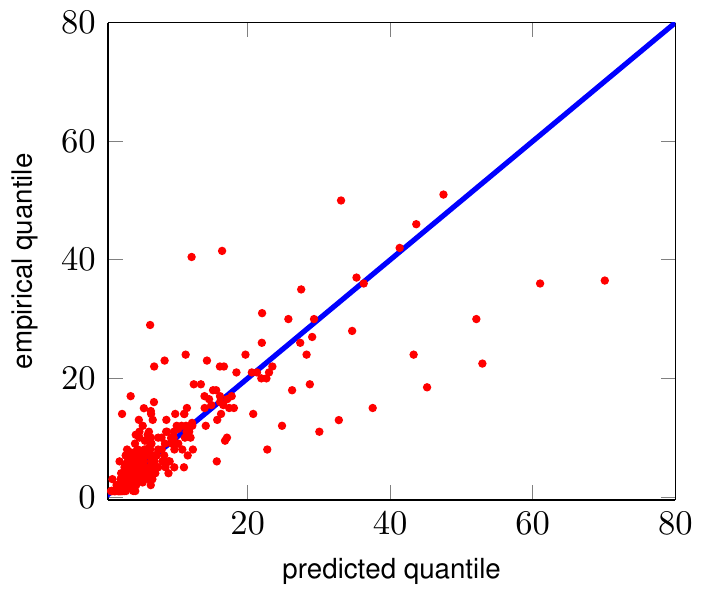}
\caption{Predicted versus empirical $0.50$-th quantile for groups with at least 50 publications.}
\label{fig:predreal50}
\end{minipage}
\hspace{0.1cm}
\begin{minipage}{0.4\linewidth}
\includegraphics[width=\textwidth]{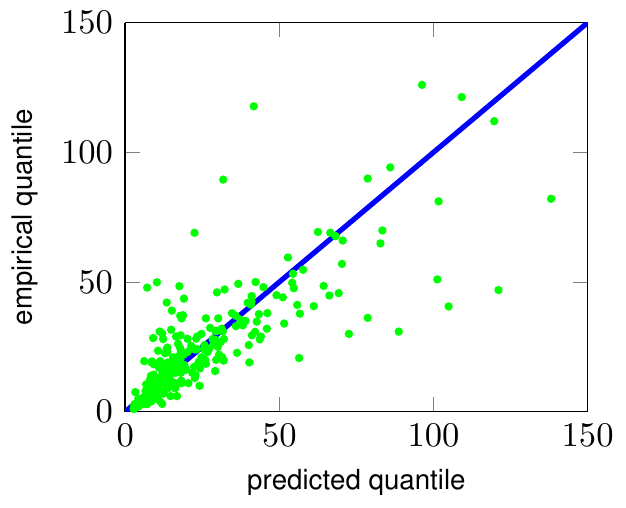}
\caption{Predicted versus empirical $0.80$-th quantile for groups with at least 50 publications.}
\label{fig:predreal80}
\end{minipage}
\hspace{0.1cm}
\begin{minipage}{0.4\linewidth}
\includegraphics[width=\textwidth]{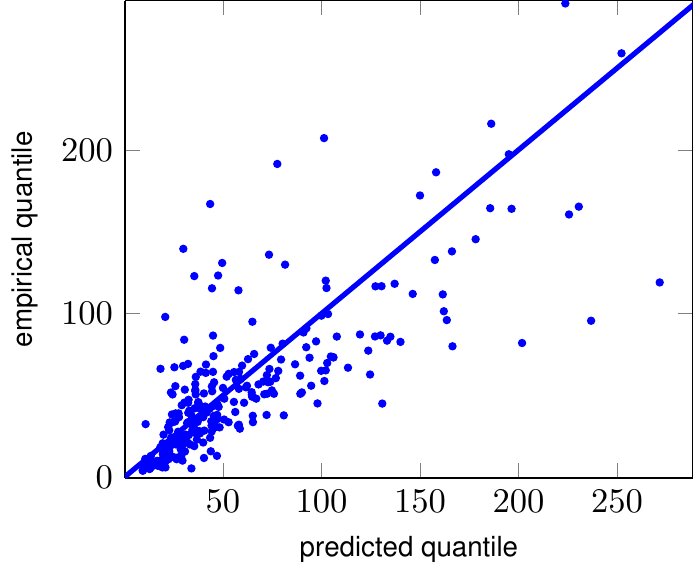}
\caption{Predicted versus empirical $0.95$-th quantile for groups with at least 50 publications.}
\label{fig:predreal95}
\end{minipage}
\hspace{0.1cm}
\begin{minipage}{0.4\linewidth}
\includegraphics[width=\textwidth]{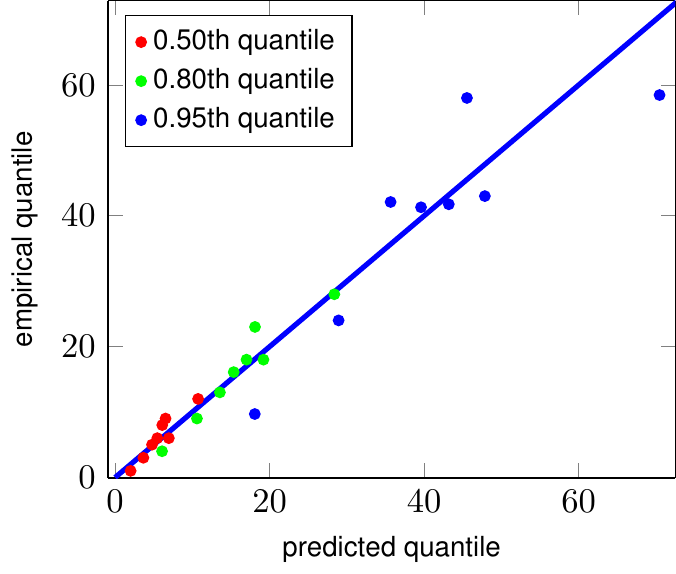}
\caption{Predicted versus empirical $0.50$-th, $0.80$-th, and $0.95$-th quantile for groups with at least 500 publications.}
\label{fig:predreallarge}
\end{minipage}
\end{figure}

To illustrate the fit of the full model, we plot the predicted quantiles against the empirical quantiles. To do so, we first create groups of publications. A group consists of publications that all have the same $c_1$ and the same impact factor, where impact factors have been rounded to halves. Figures~\ref{fig:predreal50},~\ref{fig:predreal80} and~\ref{fig:predreal95} show all groups that include at least 50 publications. The figures relate to, respectively, the $0.50$-th, $0.80$-th, and $0.95$-th quantile. Each dot in the figures corresponds to a group of publications with the same $IF$ and $c_1$. The 45-degree lines are shown as a reference. In the case of a perfect fit, all dots should be located on the 45-degree lines. We see that for many groups of publications predictions are quite accurate, but there are also quite some groups for which there is a large difference between the predicted and the empirical quantile. Taking a closer look at the data, we see that, naturally, a better fit is obtained for larger groups. In Figure~\ref{fig:predreallarge}, the predicted quantiles are again plotted against the empirical quantiles, but only groups including at least 500 publications are shown. The $0.50$-th, $0.80$-th, and $0.95$-th quantiles are presented in the same plot. For the $0.50$-th and $0.80$-th quantile, the fit is excellent. For the $0.95$-th quantile, there are more outliers. We note that estimates for high quantiles will be explored further using quantile estimators for Pareto tails in Section~\ref{sec:Parest}.

\subsection{Comparing the fit of the different models}\label{sec:comp}

\begin{figure}[htbp]
\centering
\begin{subfigure}[b]{0.3\textwidth}
\includegraphics[width=\textwidth]{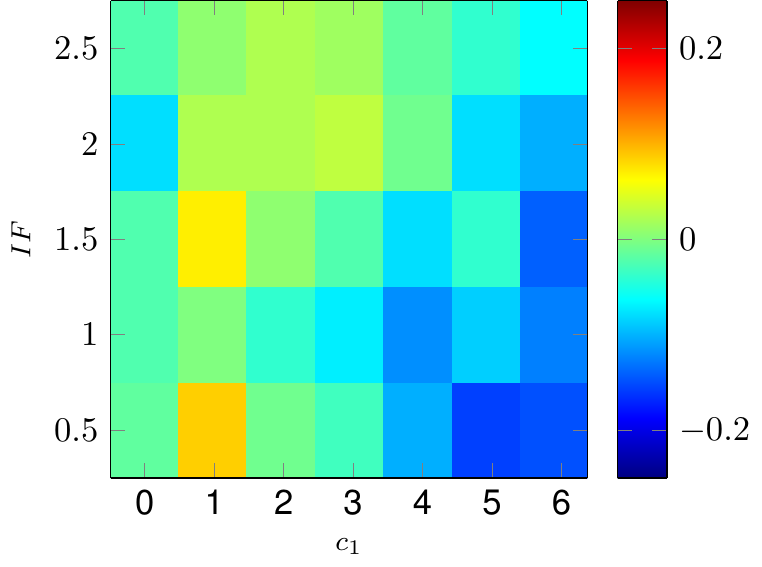}
\caption{Full model.}
\label{fig:nlessfull}
\end{subfigure}
\begin{subfigure}[b]{0.3\textwidth}
\includegraphics[width=\textwidth]{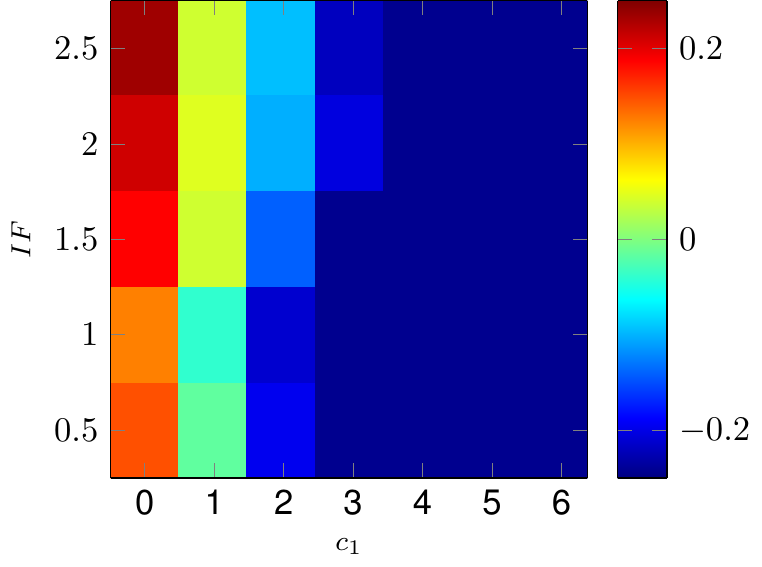}
\caption{Model with only $IF$.}
\label{fig:nlessIF}
\end{subfigure}
\begin{subfigure}[b]{0.3\linewidth}
\includegraphics[width=\textwidth]{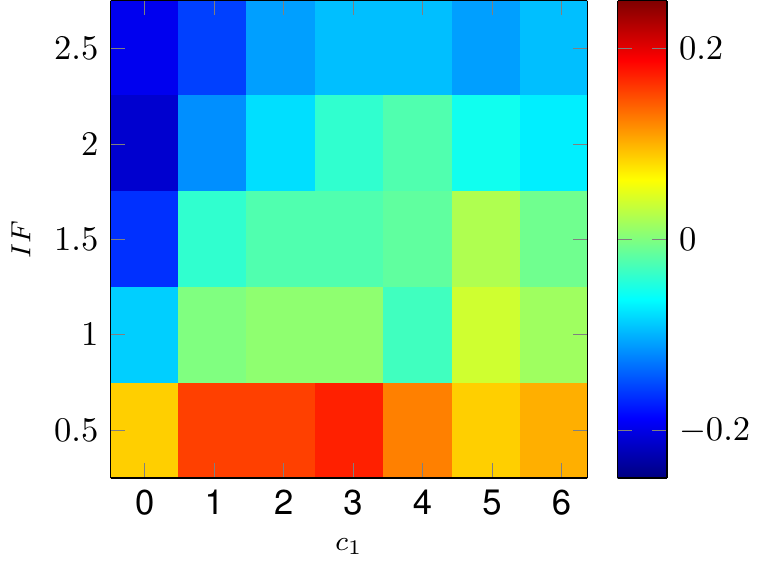}
\caption{Model with only $c_1$.}
\label{fig:nlessc1}
\end{subfigure}
\caption{Let $f$ be the fraction of publications with fewer citations than the predicted $0.50$-th quantile. The figures show $f-0.5$ for the three models for different $c_1$ and $IF$.}
\end{figure}

Figures~\ref{fig:nlessfull},~\ref{fig:nlessIF}, and~\ref{fig:nlessc1} illustrate the difference in the fit of the three models. Let $f$ be the fraction of publications with fewer citations than the predicted $0.50$-th quantile. For different groups of publications, the figures show $f-0.5$. If the model predicts correctly, we expect this value to be close to zero. This is indicated by a green color in the figures. Red colors represent positive values, indicating that the model overestimates the $0.50$-th quantile for a group of publications. Blue colors correspond to negative values, which means underestimation of the $0.50$-th quantile. Each rectangle in the figures represents a group of publications with the same impact factor and the same number of citations in the first year. For example, for the model that uses only the impact factor, the model overestimates the $0.50$-th quantile for publications with an impact factor of 2.5 and with 0 citations in the first year. It underestimates the $0.50$-th quantile for publications with an impact factor of 2.5 and with 5 citations in the first year.

Based on these figures, we see that the model which uses only the impact factor does not predict very well for publications with either a small or a large number of citations in the first year. Likewise, the model that uses only the number of early citations does not predict very well for publications with either a low or a high impact factor. Similar figures can be created for other quantiles instead of the $0.50$-th quantile. From these figures we conclude that the full model yields more accurate predictions than the other two models. This means that both the impact factor and the number of citations in the first year provide important information for predictive purposes, and that impact factors and early citations should therefore be used together to obtain accurate predictions. For this reason, in the remainder of the results, the full model is used.

\subsection{Predicting the conditional citation distribution}\label{sec:distribution}

\begin{figure}[htbp]
\centering
\begin{subfigure}[b]{0.45\textwidth}
\includegraphics[width=\textwidth]{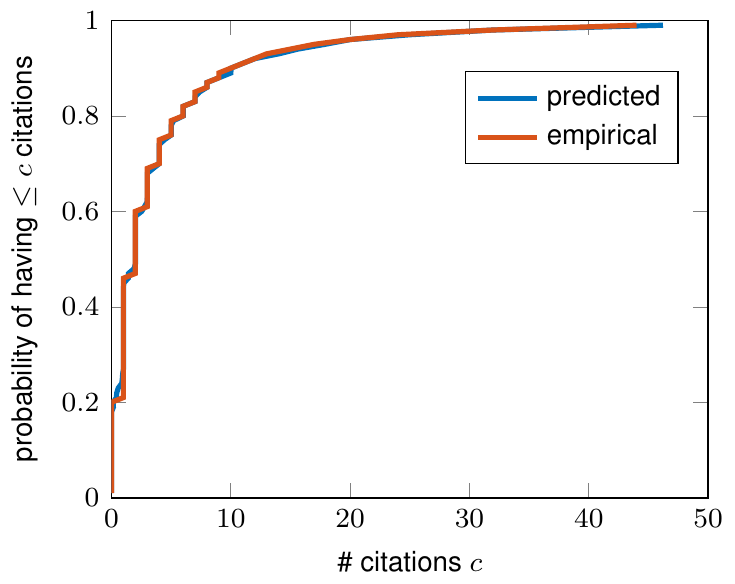}
\caption{$IF=0,c_1=0$}
\label{fig:cum00}
\end{subfigure}
\begin{subfigure}[b]{0.45\textwidth}
\includegraphics[width=\textwidth]{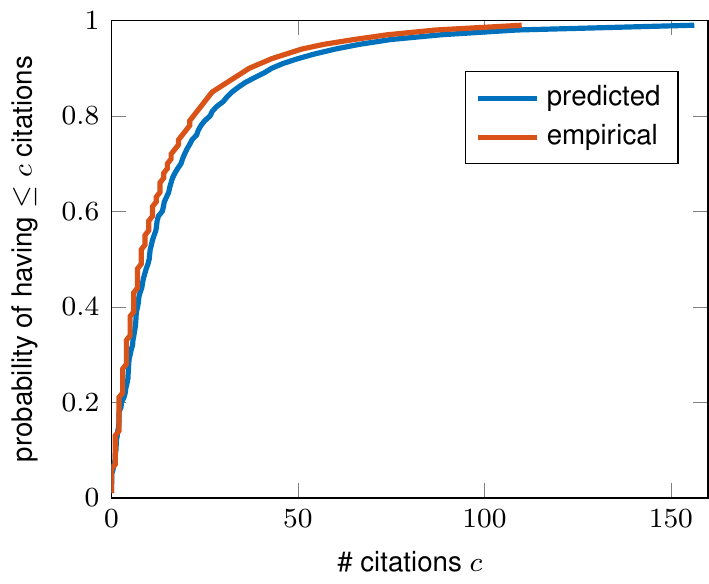}
\caption{$IF=1,c_1=1$}
\label{fig:cum11}
\end{subfigure}
\caption{Empirical and predicted conditional distribution function of the number of citations after 30 years.}
\end{figure}

Using the quantile regression model, we can predict the entire conditional distribution of the number of citations. Figures~\ref{fig:cum00} and~\ref{fig:cum11} show the predicted and empirical conditional distribution for publications that have impact factor zero and zero citations in the first year and for publications that have impact factor one and one citation in the first year respectively. It is clear that publications with the same impact factor and the same number of citations in the first year, may have different citation numbers after 30 years. For this reason, predicting the entire conditional distribution is more valuable than giving a point estimate on the number of citations that publications receive. Furthermore, the conditional distributions in the two figures are different, which indicates that it is important to take into account the influence of the impact factor and the number of early citations. The quantile regression method predicts the conditional distribution quite accurately, especially for the publications with an impact factor of zero, and zero citations in the first year.

\subsection{Predictions for later publications}\label{sec:perf}

In this section, we test whether the model fitted based on older publications also predicts well when applied to more recent publications. To this end, we first estimate the quantile regression coefficients for publications in the field of physics published in 1990. The model is fitted to predict the quantiles of the number of citations that these publications have received by the end of 2000. We then use the resulting model to predict the quantiles of the number of citations that publications in the field of physics published in 2000 have received by the end of 2010.

\begin{figure}[htbp]
\centering
\begin{minipage}{0.4\linewidth}
\centering
\includegraphics[width=\textwidth]{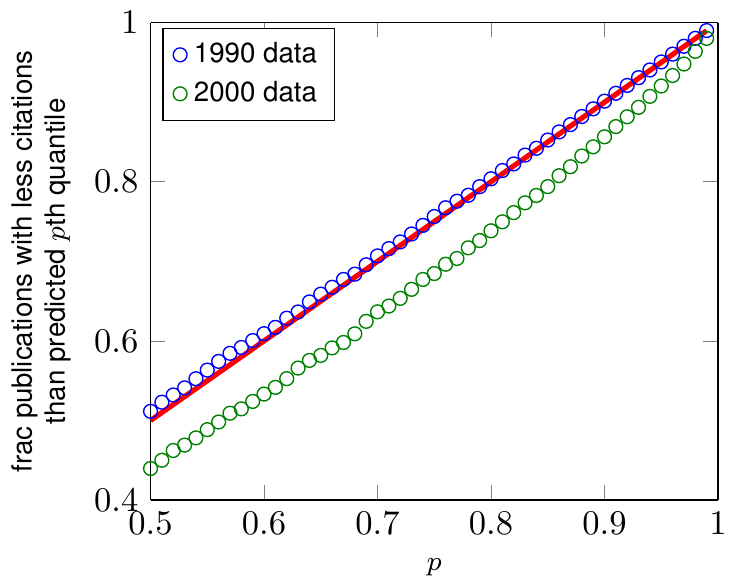}
\caption{$p$ versus the fraction of publications with fewer citations than their predicted $p$-th quantile.}
\label{fig:2000data}
\end{minipage}
\hspace{0.2cm}
\begin{minipage}{0.4\linewidth}
\centering
\includegraphics[width=\textwidth]{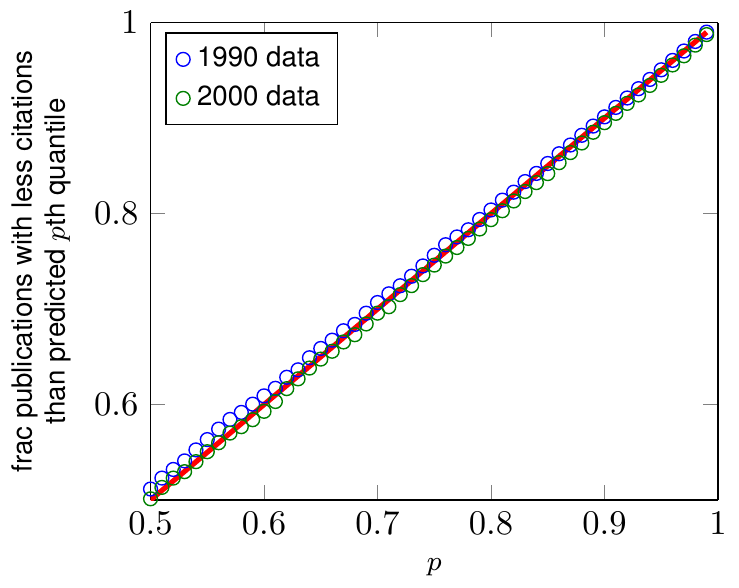}
\caption{$p$ versus the fraction of publications with fewer citations than their predicted $p$-th quantile. An adjustment has been made for increasing citation counts over time.}
\label{fig:2000adj}
\end{minipage}
\end{figure}

The predictive performance of the model is illustrated in Figure~\ref{fig:2000data}. We compute the fraction of publications that have received fewer citations than their predicted $p$-th quantile. If the model predicts well, this fraction should be $p$. In Figure~\ref{fig:2000data}, $p$ is plotted against the fraction of publications with fewer citations than their predicted $p$-th quantile. Results are shown both for publications from 1990 (which were used to fit the model) and for publications from 2000 (which were not used in model fitting). The 45-degree line is included as a reference. In the case of a perfect fit, all dots should be located on the 45-degree line.

For publications from 1990, the quantiles are predicted almost perfectly, which is to be expected because we use quantile regression (see Section~\ref{sec:qregression}). However, when the model is applied to publications from 2000, we see that the quantiles are underestimated. For example, only around 43\% of the publications from 2000 have received fewer citations than their predicted $0.50$-th quantile. This disappointing result must be due to structural changes that have taken place over time and that cause a model fitted to older data not to perform well when applied to newer data. In particular, there is a trend to include more and more references in publications, and as a result of this trend, the average number of citations that publications receive has increased over time~\cite{wallace2009}. Because the model is fitted based on older publications, which have lower citation counts than more recent publications, the model underestimates the quantiles for more recent publications.

We want to adjust the predictions of the model for the increase over time in the average number of citations per publication. To do so, we make predictions based on normalized data. This means that all inputs and outputs of the model are divided by their average value. For example, the number of citations of a publication in the first year, $c_1$, is divided by the average value of $c_1$ over all publications. Similarly, the number of citations of a publication after 10 years is divided by the average number of citations after 10 years over all publications. The quantile regression model is fitted on the normalized data from 1990. The data from 2000 are normalized in the same way, and the model fitted based on publications from 1990 is used to predict the quantiles for publications from 2000. The resulting predictions are normalized predictions with respect to the average number of citations after 10 years over all publications from 2000. Hence, a predicted quantile of for example 2 means that the quantile equals twice the average number of citations after 10 years over all publications from 2000.

Figure~\ref{fig:2000adj} shows that the performance of the model for publications from 2000 has become just as good as for publications from 1990. Like in Figure~\ref{fig:2000data}, $p$ is plotted against the fraction of publications with fewer citations than their predicted $p$-th quantile. The good performance of the model based on normalized data indicates that the increasing number of citations received by publications is indeed responsible for the underestimation that can be observed in Figure~\ref{fig:2000data}. The use of normalized data solves this problem.


\section{Tail quantiles}\label{sec:Parest}

In this section, we take a closer look at high quantiles using a quantile estimation technique from extreme value theory.

\subsection{Tail of the citation distribution}\label{sec:zenga}

In the literature, a lot of attention has been paid to the tail of citation distributions, and in particular to the question whether this tail follows a Pareto distribution. The possibility of a Pareto tail was already suggested by~\citeA{price1976}.~\citeA{redner1998} analyzed a large data set of publications and their citations and observed that the tail of the distribution of citations over publications can be described by a power law.~\citeA{clauset2009} proposed a statistical methodology for testing the presence of power-law behavior in empirical data. Based on the same data set as~\citeA{redner1998}, they concluded that for citation distributions a power-law tail cannot be ruled out. The methodology of~\citeA{clauset2009} was also used by~\citeA{albarran2011skewness}, who found that in a large number of scientific fields citation distributions seem to have a power-law tail.

Formally, let $X$ be the number of future citations. The random variable $X$ has a Pareto tail if for some $x_l$ we have
\begin{equation*}
P(X>x)=wx^{-\alpha},\quad\quad x\geq x_l.
\end{equation*}
Here $w$ and $\alpha$ are parameters. $\alpha$ is also called the tail exponent.

There are many ways to test whether a distribution has a Pareto tail. Here we use the Zenga plot proposed by~\citeA{cirillo2013}. The motivation behind this method is that it allows distinguishing between a Pareto tail and a lognormal tail, while many other methods, such as the QQ-plot, fail to detect this difference. 

Let $F$ be the distribution function of a random variable $X$. The Zenga curve $Z$ is defined as
\begin{eqnarray*}
Z(u)=&1-\frac{Q^-(u)}{Q^+(u)},\quad &0<u<1,\\
Q^-(u)=&\frac{1}{u}\int_0^uF^{-1}(s)ds,\quad &0\leq u\leq1,\\
Q^+(u)=&\frac{1}{1-u}\int_u^1F^{-1}(s)ds,\quad &0\leq u\leq1.
\end{eqnarray*}
Hence, the Zenga curve is a measure of how much weight of the distribution lies below the $u$-th quantile relative to how much weight lies above the $u$-th quantile, as a function of $u$. The Zenga curve has different shapes for different distributions. For the lognormal distribution, the Zenga curve is a straight line, while for Pareto distributions, the Zenga curve is a convex increasing function~\cite{cirillo2013}. 

\begin{figure}[ht]
\centering
\begin{minipage}{0.4\linewidth}
\includegraphics[width=\textwidth]{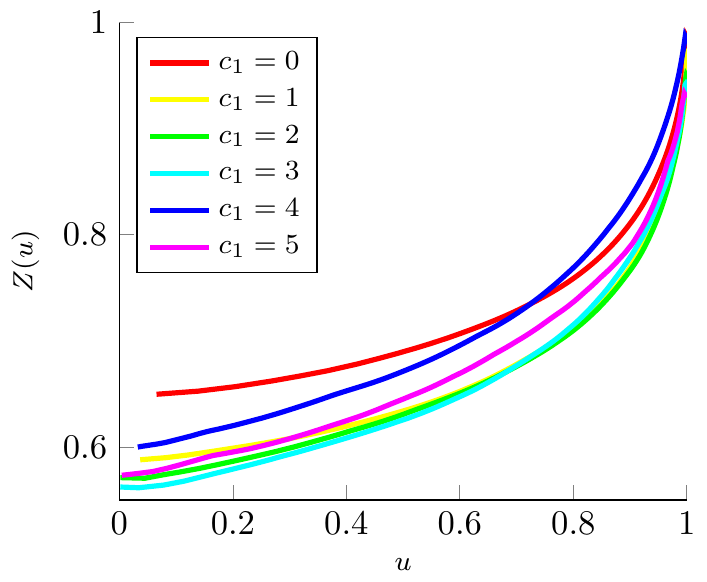}
\caption{Zenga plot for different $c_1$.}
\label{fig:zengac}
\end{minipage}
\hspace{0.1cm}
\begin{minipage}{0.4\linewidth}
\includegraphics[width=\textwidth]{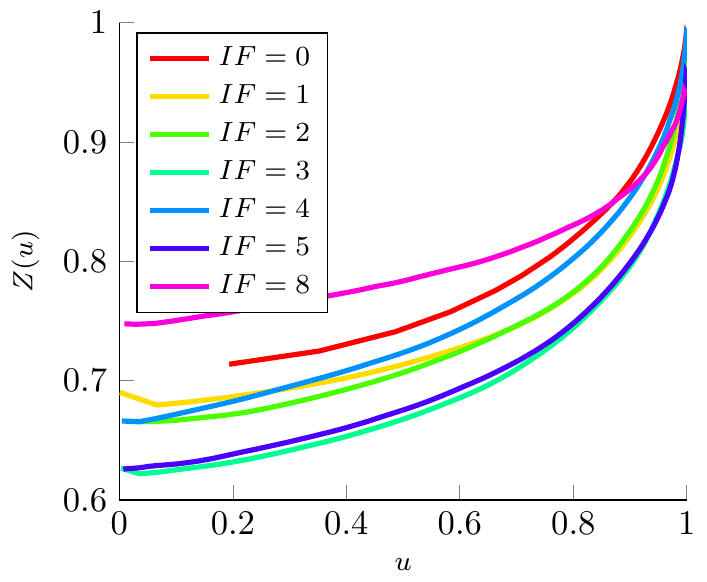}
\caption{Zenga plot for different $IF$.}
\label{fig:zengaimp}
\end{minipage}
\end{figure}

Figures~\ref{fig:zengac} and~\ref{fig:zengaimp} show the empirical Zenga plot for groups of publications with either the same number of citations in the first year or the same impact factor. The Zenga curves are clearly convex increasing functions. This indicates that for all groups of publications the citation distribution has a Pareto tail.

\subsection{Pareto quantile estimator}\label{sec:parest}

Under the assumption of a Pareto tail, we can estimate high quantiles using the estimator proposed by~\citeA{dekkers1989}. For a group of publications with given $IF$ and $c_1$, this estimator is of the form
\begin{equation}\label{eq:Paretoest}
q(p|IF,c_1)=X_{(n-k,n)}\left(\frac{k}{n(1-p)}\right)^{1/\alpha}.
\end{equation}
Here $n:=n(IF,c_1)$ is the number of publications in the group of publications with given $IF$, and $c_1$ and $X_{(n-k,n)}:=X_{(n,n-k)}(IF,c_1)$ is the number of citations of the $k$-th most cited publication in this group. Furthermore, $k:=k(IF,c_1)$ is the threshold where the Pareto tail starts. So the $k$ publications with the largest number of citations follow a Pareto distribution.

The Pareto tail starts at the $\left(1-\frac{k}{n}\right)$-th quantile. This threshold quantile will be called the $p^*$-th quantile. Thus, $X_{(n-k,n)}$ is the empirical value of the $p^*$-th quantile. The tail index $\alpha$ is estimated using the Hill estimator~\cite{hill1975}. The threshold $k$ and hence the threshold quantile $p^*$ is estimated using the procedure suggested by~\cite{beirlant2007}. This procedure minimizes an approximation of the asymptotic mean squared error of the estimate of $\alpha$. The resulting estimate for $p^*$ is $p^*=0.95$.

Figure~\ref{fig:Parreal} plots the quantiles that are predicted by Eq.~\eqref{eq:Paretoest} against the empirical quantiles, when $X_{(n-k,n)}$ is given. Note that the accuracy of the predictions is inconclusive because we rely on $X_{(n-k,n)}$, which is not known in practice. However, we see that Eq.~\eqref{eq:Paretoest} captures the behavior of high quantiles quite accurately. In the next section, we will obtain insight into the predictions for high quantiles by linking the Pareto estimator to the quantile regression estimator.

%
\begin{figure}[htbp]

\end{figure}

\subsection{Linking the Pareto and regression estimators}

As mentioned above, the problem with Eq.~\eqref{eq:Paretoest} is that it uses $X_{(n-k,n)}$, the empirical value of the ${p^*}$-th quantile, at which the Pareto tail starts. In practice, $X_{(n-k,n)}$ is not known and needs to be predicted. A natural way to overcome this is to replace $X_{(n-k,n)}$ by the predicted ${p^*}$-th quantile from the quantile regression:
\begin{equation*}
q(p^*|IF,c_1)=\tilde{C}_{p^*}IF^{\beta_{p^*}}(c_1+k_0)^{\gamma_{p^*}}.
\end{equation*}
We then obtain the following estimator for the tail quantiles:
\begin{equation}\label{eq:paretofull}
q(p|IF,c_1)=\left(\frac{1-p^*}{1-p}\right)^{1/\alpha}\tilde{C}_{p^*}IF^{\beta_{p^*}}({c_1}+k_0)^{\gamma_{p^*}},\quad\quad p\geq p^*,
\end{equation}
where we used the identity $\frac{k}{n}=1-p^*$.

\begin{figure}[htbp]
\centering
\centering
\begin{minipage}{0.4\linewidth}
\centering
\includegraphics[width=\textwidth]{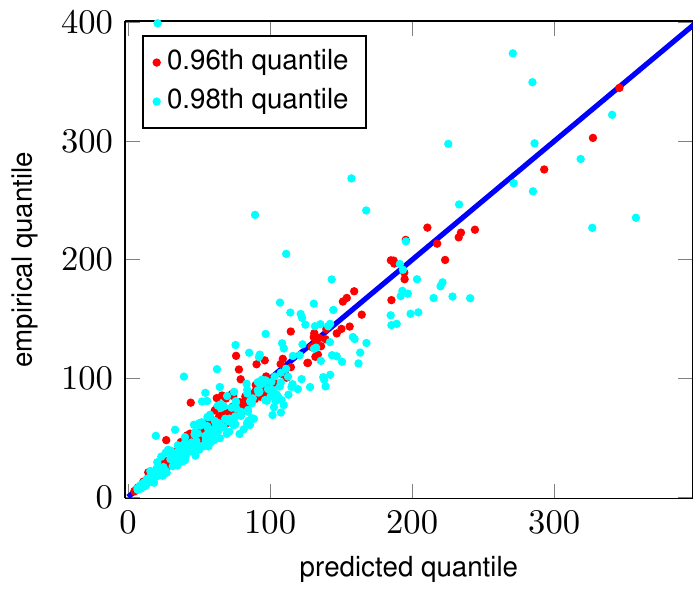}
\caption{Predicted $p$-th quantile obtained from the Pareto estimator versus empirical $p$-th quantile for $p=0.96$ and $p=0.98$ and for groups with at least 50 publications.}
\label{fig:Parreal}
\end{minipage}
\hspace{0.1cm}
\begin{minipage}{0.4\linewidth}
\centering
\includegraphics[width=\textwidth]{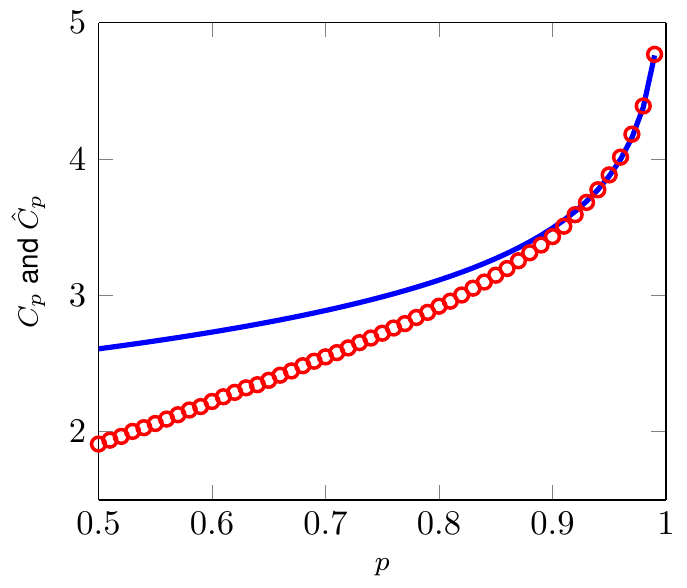}
\caption{$\hat C_p$ from Eq.~\eqref{eq:ccheck} (blue line) and $C_p$ from regression (red dots) for $p$-th quantile.}
\label{fig:ccheck}
\end{minipage}
\end{figure}

We can now explain the behavior of $C_p=\ln(\tilde{C_p})$ in Figure~\ref{fig:quc} by comparing Eq.~\eqref{eq:paretofull} to the regression estimator. If the two estimators were equal, then for $p\ge p^*$ we would have
\begin{equation*}
\left(\frac{1-p^*}{1-p}\right)^{1/\alpha}\tilde{C}_{p^*}IF^{\beta_{p^*}}({c_1}+k_0)^{\gamma_{p^*}}=\tilde{C}_{p}IF^{\beta_{p}}({c_1}+k_0)^{\gamma_{p}}.
\end{equation*}
Hence, if we assume that $\beta_p$ and $\gamma_p$ are constant for large values of $p$, this suggests that we may use $\left(\frac{1-p^*}{1-p}\right)^{1/\alpha}\tilde{C}_{p^*}$ for $\tilde{C}_p$. By taking the logarithm, we obtain the following proxy $\hat C_p$ for the regression coefficient $C_p$:
\begin{equation}\label{eq:ccheck}
\hat C_p=C_{p^*}+\frac{1}{\alpha}\left(\ln(1-p^*)-\ln(1-p)\right), \quad\quad p\geq p^*.
\end{equation}
In Figure~\ref{fig:ccheck}, $C_p$ and $\hat C_p$ are plotted. The blue line corresponds to $\hat C_p$ given by Eq.~\eqref{eq:ccheck} with $p^*=0.95$. For completeness, we plot this line for all $p\in[0.5,0.99]$. The red dots correspond to the regression coefficients $C_p$. We see that Eq.~\eqref{eq:ccheck} indeed can be used as an analytical description of $C_p$ when $p\ge p^*$. For $p\ge 0.95$, there is an excellent agreement between $C_p$ and $\hat C_p$. In fact, Eq.~\eqref{eq:ccheck} shows a good agreement for all $p\ge 0.9$.

\section{Sensitivity of parameters to the field of science}\label{ref:sensitivity}

\begin{figure}[htbp]
\centering
\begin{subfigure}[b]{0.32\textwidth}
\includegraphics[width=\textwidth]{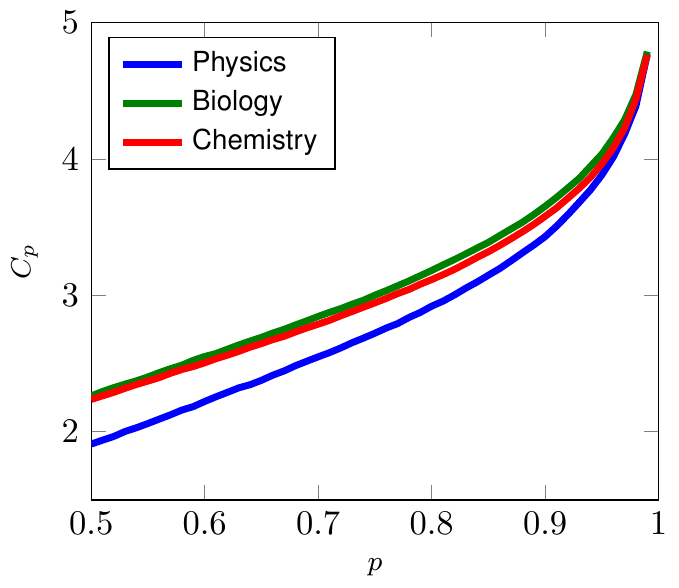}
\caption{$C_p$}
\label{fig:Cfield}
\end{subfigure}
\begin{subfigure}[b]{0.32\textwidth}
\includegraphics[width=\textwidth]{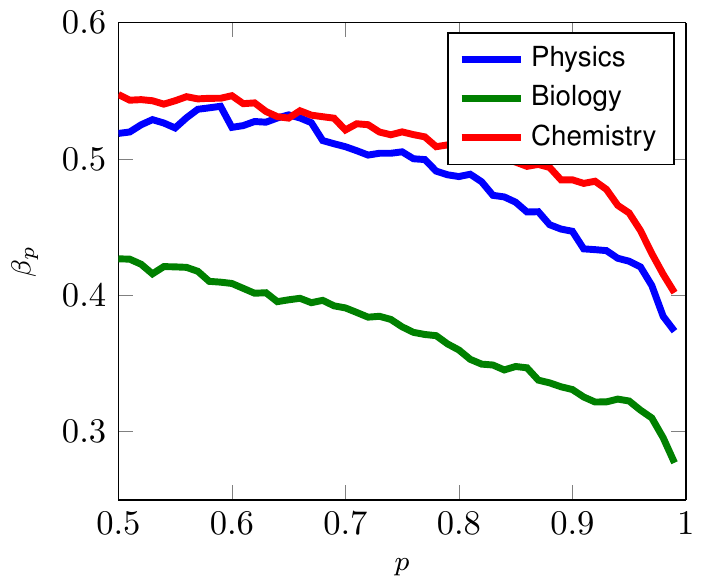}
\caption{$\beta_p$}
\label{fig:betafield}
\end{subfigure}
\begin{subfigure}[b]{0.32\textwidth}
\includegraphics[width=\textwidth]{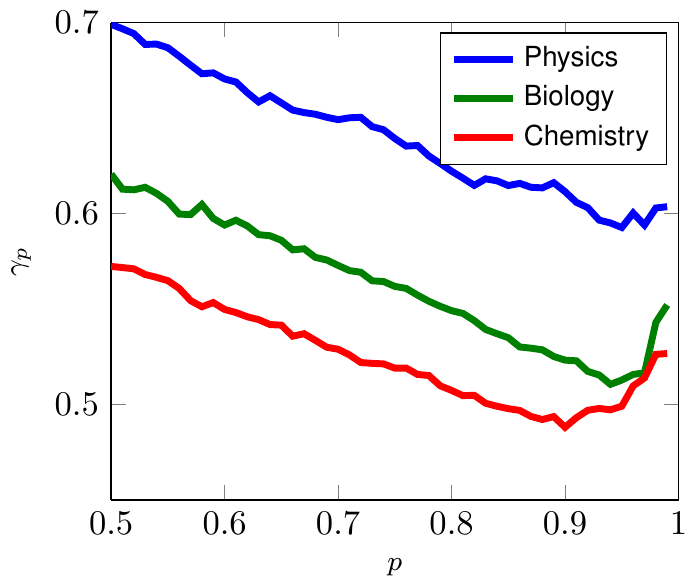}
\caption{$\gamma_p$}
\label{fig:gammafield}
\end{subfigure}
\caption{Quantile regression coefficients for $p$-th quantile versus $p$ for three fields of science.}
\end{figure}

In the previous sections, data from the field of physics was used to fit the coefficients of the regression model. In this section, we study the influence of the field of science on the regression coefficients. We again consider publications published in 1984, and the quantile regression is again used to predict the conditional quantiles of the distribution of the number of citations received by these publications by the end of 2013. The coefficients $C_p$, $\beta_p$, and $\gamma_p$ obtained from the quantile regression are plotted in Figures~\ref{fig:Cfield},~\ref{fig:betafield} and~\ref{fig:gammafield} for three different fields: Biology, chemistry, and physics.

The regression coefficient $C_p$ is higher for publications in biology and chemistry than for publications in physics. Other things being equal, this means that publications in biology and chemistry receive more citations than publications in physics. This is the case mainly for the lower quantiles. For the higher quantiles, the differences in the coefficient $C_p$ are small. The regression coefficient $\beta_p$ is lower for biology publications than for publications in physics and chemistry. This means that in biology the impact factor is less determining for the long-term citation impact of publications. However, the differences in this coefficient are small. The coefficient $\gamma_p$ is highest for publications in physics. Hence, in physics the number of citations that a publication has received in the first year is more determining for the publication's long-term citation impact than in biology and chemistry.


\section{Conclusions}\label{sec:conc}

We have proposed a model to predict a probability distribution for the future number of citations of a publication. Two predictors are considered in the model: The impact factor of the journal in which a publication has appeared and the number of citations received by a publication in the first year after its appearance. The proposed model is based on quantile regression. The good fit of the model indicates that quantile regression is a suitable tool to predict the quantiles of the probability distribution of a publication's future number of citations. We have found that the quantile regression coefficients $\beta_p$ and $\gamma_p$, corresponding to respectively the impact factor and the number of early citations, are not stable in the quantile $p$. Hence, the influence of the impact factor and the number of early citations on the long-term citation impact of a publication is different for different quantiles.

Three variants of our prediction model have been studied. The variant in which both the impact factor and the number of early citations are used turns out to fit the data better than the variants in which only one of the two predictors is included. This means that both the impact factor and the number of early citations are important to predict the probability distribution of a publication's future number of citations.

Importantly, our proposed model provides accurate predictions also for publications that were published later than the publications used for estimating the model parameters. However, in order to obtain these accurate predictions, it is necessary to normalize all inputs and outputs of the model by their average value.

We have also investigated the tail of the citation distributions obtained in our analysis. Zenga plots~\cite{cirillo2013} have been used for this purpose. It turns out that the tail of our citation distributions can be approximated by a Pareto distribution. Using an estimator for the tail quantiles of a Pareto distribution~\cite{dekkers1989}, we have obtained an explicit equation for the regression coefficient $C_p$ in our model for high quantiles $p$.

There are a number of issues that require further research. First of all, further research may focus on the fitness factor that we use in our model. Following~\citeA{ke2013}, we have assumed that the fitness factor is a product of our two predictors each raised to a certain power. Other ways of modeling the fitness factor may also be investigated.

The analysis presented in this paper is based on publications in the field of physics. This is a broad field consisting of many different subfields. These subfields probably all have their own citation practices. Differences in citation practices between fields or subfields have not been taken into account in our prediction approach. Further research may focus on linking our prediction approach to the literature on field normalization of citation-based indicators.

Another issue for further research is the use of other predictors, in addition to impact factor and early citations. In Section~\ref{sec:lit}, we already suggested some possibilities: Number of downloads of a publication, number of readers according to a service such as Mendeley, and other types of altmetric indicators. Further research may investigate the effect of adding these predictors to our model. In particular, it would be interesting to find out whether the use of additional predictors decreases the level of uncertainty in predictions of long-term citation impact.

Finally, perhaps the most challenging issue for further research is to make predictions of long-term citation impact not only for individual publications but also for the entire publication oeuvre of a researcher, a research group, or a research institution~\cite<e.g.,>{acuna2012future,bornmann2013problem}. Moving from predictions at the individual publication level to predictions at the level of oeuvres of publications is far from trivial. A prediction approach that yields accurate results at the individual publication level may provide biased results when it is used at the level of oeuvres of publications.

\bibliographystyle{apacite}
\bibliography{references}

\begin{thebibliography}{}

\bibitem [\protect \citeauthoryear {%
Abramo%
, D'Angelo%
\BCBL {}\ \BBA {} Di~Costa%
}{%
Abramo%
\ \protect \BOthers {.}}{%
{\protect \APACyear {2010}}%
}]{%
abramo2010citations}
\APACinsertmetastar {%
abramo2010citations}%
\begin{APACrefauthors}%
Abramo, G.%
, D'Angelo, C.%
\BCBL {}\ \BBA {} Di~Costa, F.%
\end{APACrefauthors}%
\unskip\
\newblock
\APACrefYearMonthDay{2010}{}{}.
\newblock
{\BBOQ}\APACrefatitle {Citations versus journal impact factor as proxy of
  quality: Could the latter ever be preferable?} {Citations versus journal
  impact factor as proxy of quality: Could the latter ever be
  preferable?}{\BBCQ}
\newblock
\APACjournalVolNumPages{Scientometrics}{84}{3}{821--833}.
\PrintBackRefs{\CurrentBib}

\bibitem [\protect \citeauthoryear {%
Acuna%
, Allesina%
\BCBL {}\ \BBA {} Kording%
}{%
Acuna%
\ \protect \BOthers {.}}{%
{\protect \APACyear {2012}}%
}]{%
acuna2012future}
\APACinsertmetastar {%
acuna2012future}%
\begin{APACrefauthors}%
Acuna, D.%
, Allesina, S.%
\BCBL {}\ \BBA {} Kording, K.%
\end{APACrefauthors}%
\unskip\
\newblock
\APACrefYearMonthDay{2012}{}{}.
\newblock
{\BBOQ}\APACrefatitle {Future impact: Predicting scientific success} {Future
  impact: Predicting scientific success}.{\BBCQ}
\newblock
\APACjournalVolNumPages{Nature}{489}{7415}{201--202}.
\PrintBackRefs{\CurrentBib}

\bibitem [\protect \citeauthoryear {%
Adams%
}{%
Adams%
}{%
{\protect \APACyear {2005}}%
}]{%
adams2005early}
\APACinsertmetastar {%
adams2005early}%
\begin{APACrefauthors}%
Adams, J.%
\end{APACrefauthors}%
\unskip\
\newblock
\APACrefYearMonthDay{2005}{}{}.
\newblock
{\BBOQ}\APACrefatitle {Early citation counts correlate with accumulated impact}
  {Early citation counts correlate with accumulated impact}.{\BBCQ}
\newblock
\APACjournalVolNumPages{Scientometrics}{63}{3}{567--581}.
\PrintBackRefs{\CurrentBib}

\bibitem [\protect \citeauthoryear {%
Albarr{\'a}n%
, Crespo%
, Ortu{\~n}o%
\BCBL {}\ \BBA {} Ruiz-Castillo%
}{%
Albarr{\'a}n%
\ \protect \BOthers {.}}{%
{\protect \APACyear {2011}}%
}]{%
albarran2011skewness}
\APACinsertmetastar {%
albarran2011skewness}%
\begin{APACrefauthors}%
Albarr{\'a}n, P.%
, Crespo, J.%
, Ortu{\~n}o, I.%
\BCBL {}\ \BBA {} Ruiz-Castillo, J.%
\end{APACrefauthors}%
\unskip\
\newblock
\APACrefYearMonthDay{2011}{}{}.
\newblock
{\BBOQ}\APACrefatitle {The skewness of science in 219 sub-fields and a number
  of aggregates} {The skewness of science in 219 sub-fields and a number of
  aggregates}.{\BBCQ}
\newblock
\APACjournalVolNumPages{Scientometrics}{88}{2}{385--397}.
\PrintBackRefs{\CurrentBib}

\bibitem [\protect \citeauthoryear {%
Beirlant%
, Gl{\"a}nzel%
, Carbonez%
\BCBL {}\ \BBA {} Leemans%
}{%
Beirlant%
\ \protect \BOthers {.}}{%
{\protect \APACyear {2007}}%
}]{%
beirlant2007}
\APACinsertmetastar {%
beirlant2007}%
\begin{APACrefauthors}%
Beirlant, J.%
, Gl{\"a}nzel, W.%
, Carbonez, A.%
\BCBL {}\ \BBA {} Leemans, H.%
\end{APACrefauthors}%
\unskip\
\newblock
\APACrefYearMonthDay{2007}{}{}.
\newblock
{\BBOQ}\APACrefatitle {Scoring research output using statistical quantile
  plotting} {Scoring research output using statistical quantile
  plotting}.{\BBCQ}
\newblock
\APACjournalVolNumPages{Journal of Informetrics}{1}{3}{185--192}.
\PrintBackRefs{\CurrentBib}

\bibitem [\protect \citeauthoryear {%
Bornmann%
}{%
Bornmann%
}{%
{\protect \APACyear {2013}}%
}]{%
bornmann2013problem}
\APACinsertmetastar {%
bornmann2013problem}%
\begin{APACrefauthors}%
Bornmann, L.%
\end{APACrefauthors}%
\unskip\
\newblock
\APACrefYearMonthDay{2013}{}{}.
\newblock
{\BBOQ}\APACrefatitle {The problem of citation impact assessments for recent
  publication years in institutional evaluations} {The problem of citation
  impact assessments for recent publication years in institutional
  evaluations}.{\BBCQ}
\newblock
\APACjournalVolNumPages{Journal of Informetrics}{7}{3}{722--729}.
\PrintBackRefs{\CurrentBib}

\bibitem [\protect \citeauthoryear {%
Bornmann%
, Leydesdorff%
\BCBL {}\ \BBA {} Wang%
}{%
Bornmann%
\ \protect \BOthers {.}}{%
{\protect \APACyear {2013}}%
}]{%
bornmann2013percentile}
\APACinsertmetastar {%
bornmann2013percentile}%
\begin{APACrefauthors}%
Bornmann, L.%
, Leydesdorff, L.%
\BCBL {}\ \BBA {} Wang, J.%
\end{APACrefauthors}%
\unskip\
\newblock
\APACrefYearMonthDay{2013}{}{}.
\newblock
{\BBOQ}\APACrefatitle {Which percentile-based approach should be preferred for
  calculating normalized citation impact values? An empirical comparison of
  five approaches including a newly developed citation-rank approach (P100)}
  {Which percentile-based approach should be preferred for calculating
  normalized citation impact values? an empirical comparison of five approaches
  including a newly developed citation-rank approach (p100)}.{\BBCQ}
\newblock
\APACjournalVolNumPages{Journal of Informetrics}{7}{4}{933--944}.
\PrintBackRefs{\CurrentBib}

\bibitem [\protect \citeauthoryear {%
Brody%
, Harnad%
\BCBL {}\ \BBA {} Carr%
}{%
Brody%
\ \protect \BOthers {.}}{%
{\protect \APACyear {2006}}%
}]{%
brody2006earlier}
\APACinsertmetastar {%
brody2006earlier}%
\begin{APACrefauthors}%
Brody, T.%
, Harnad, S.%
\BCBL {}\ \BBA {} Carr, L.%
\end{APACrefauthors}%
\unskip\
\newblock
\APACrefYearMonthDay{2006}{}{}.
\newblock
{\BBOQ}\APACrefatitle {Earlier web usage statistics as predictors of later
  citation impact} {Earlier web usage statistics as predictors of later
  citation impact}.{\BBCQ}
\newblock
\APACjournalVolNumPages{Journal of the American Society for Information Science
  and Technology}{57}{8}{1060--1072}.
\PrintBackRefs{\CurrentBib}

\bibitem [\protect \citeauthoryear {%
Burrell%
}{%
Burrell%
}{%
{\protect \APACyear {2003}}%
}]{%
burrell2003predicting}
\APACinsertmetastar {%
burrell2003predicting}%
\begin{APACrefauthors}%
Burrell, Q.%
\end{APACrefauthors}%
\unskip\
\newblock
\APACrefYearMonthDay{2003}{}{}.
\newblock
{\BBOQ}\APACrefatitle {Predicting future citation behavior} {Predicting future
  citation behavior}.{\BBCQ}
\newblock
\APACjournalVolNumPages{Journal of the American Society for Information Science
  and Technology}{54}{5}{372--378}.
\PrintBackRefs{\CurrentBib}

\bibitem [\protect \citeauthoryear {%
Cirillo%
}{%
Cirillo%
}{%
{\protect \APACyear {2013}}%
}]{%
cirillo2013}
\APACinsertmetastar {%
cirillo2013}%
\begin{APACrefauthors}%
Cirillo, P.%
\end{APACrefauthors}%
\unskip\
\newblock
\APACrefYearMonthDay{2013}{}{}.
\newblock
{\BBOQ}\APACrefatitle {Are your data really Pareto distributed?} {Are your data
  really pareto distributed?}{\BBCQ}
\newblock
\APACjournalVolNumPages{Physica A}{392}{23}{5947--5962}.
\PrintBackRefs{\CurrentBib}

\bibitem [\protect \citeauthoryear {%
Clauset%
, Shalizi%
\BCBL {}\ \BBA {} Newman%
}{%
Clauset%
\ \protect \BOthers {.}}{%
{\protect \APACyear {2009}}%
}]{%
clauset2009}
\APACinsertmetastar {%
clauset2009}%
\begin{APACrefauthors}%
Clauset, A.%
, Shalizi, C.%
\BCBL {}\ \BBA {} Newman, M.%
\end{APACrefauthors}%
\unskip\
\newblock
\APACrefYearMonthDay{2009}{}{}.
\newblock
{\BBOQ}\APACrefatitle {Power-law distributions in empirical data} {Power-law
  distributions in empirical data}.{\BBCQ}
\newblock
\APACjournalVolNumPages{SIAM Review}{51}{4}{661--703}.
\PrintBackRefs{\CurrentBib}

\bibitem [\protect \citeauthoryear {%
Costas%
, Zahedi%
\BCBL {}\ \BBA {} Wouters%
}{%
Costas%
\ \protect \BOthers {.}}{%
{\protect \APACyear {{\protect \BIP {}}}}%
}]{%
costas_inpress_altmetrics}
\APACinsertmetastar {%
costas_inpress_altmetrics}%
\begin{APACrefauthors}%
Costas, R.%
, Zahedi, Z.%
\BCBL {}\ \BBA {} Wouters, P.%
\end{APACrefauthors}%
\unskip\
\newblock
\APACrefYearMonthDay{{\protect \BIP {}}}{}{}.
\newblock
{\BBOQ}\APACrefatitle {Do ``altmetrics'' correlate with citations? Extensive
  comparison of altmetric indicators with citations from a multidisciplinary
  perspective} {Do ``altmetrics'' correlate with citations? extensive
  comparison of altmetric indicators with citations from a multidisciplinary
  perspective}.{\BBCQ}
\newblock
\APACjournalVolNumPages{Journal of the Association for Information Science and
  Technology}{}{}{}.
\PrintBackRefs{\CurrentBib}

\bibitem [\protect \citeauthoryear {%
Dekkers%
, Einmahl%
\BCBL {}\ \BBA {} De~Haan%
}{%
Dekkers%
\ \protect \BOthers {.}}{%
{\protect \APACyear {1989}}%
}]{%
dekkers1989}
\APACinsertmetastar {%
dekkers1989}%
\begin{APACrefauthors}%
Dekkers, A\BPBI L.%
, Einmahl, J\BPBI H.%
\BCBL {}\ \BBA {} De~Haan, L.%
\end{APACrefauthors}%
\unskip\
\newblock
\APACrefYearMonthDay{1989}{}{}.
\newblock
{\BBOQ}\APACrefatitle {A moment estimator for the index of an extreme-value
  distribution} {A moment estimator for the index of an extreme-value
  distribution}.{\BBCQ}
\newblock
\APACjournalVolNumPages{Annals of Statistics}{17}{4}{1833--1855}.
\PrintBackRefs{\CurrentBib}

\bibitem [\protect \citeauthoryear {%
De~Solla~Price%
}{%
De~Solla~Price%
}{%
{\protect \APACyear {1976}}%
}]{%
price1976}
\APACinsertmetastar {%
price1976}%
\begin{APACrefauthors}%
De~Solla~Price, D\BPBI J.%
\end{APACrefauthors}%
\unskip\
\newblock
\APACrefYearMonthDay{1976}{}{}.
\newblock
{\BBOQ}\APACrefatitle {A general theory of bibliometric and other cumulative
  advantage processes} {A general theory of bibliometric and other cumulative
  advantage processes}.{\BBCQ}
\newblock
\APACjournalVolNumPages{Journal of the American Society for Information
  Science}{27}{5}{292--306}.
\PrintBackRefs{\CurrentBib}

\bibitem [\protect \citeauthoryear {%
Didegah%
\ \BBA {} Thelwall%
}{%
Didegah%
\ \BBA {} Thelwall%
}{%
{\protect \APACyear {2013}}%
{\protect \APACexlab {{\protect \BCnt {1}}}}}]{%
didegah2013determinants}
\APACinsertmetastar {%
didegah2013determinants}%
\begin{APACrefauthors}%
Didegah, F.%
\BCBT {}\ \BBA {} Thelwall, M.%
\end{APACrefauthors}%
\unskip\
\newblock
\APACrefYearMonthDay{2013{\protect \BCnt {1}}}{}{}.
\newblock
{\BBOQ}\APACrefatitle {Determinants of research citation impact in nanoscience
  and nanotechnology} {Determinants of research citation impact in nanoscience
  and nanotechnology}.{\BBCQ}
\newblock
\APACjournalVolNumPages{Journal of the American Society for Information Science
  and Technology}{64}{5}{1055--1064}.
\PrintBackRefs{\CurrentBib}

\bibitem [\protect \citeauthoryear {%
Didegah%
\ \BBA {} Thelwall%
}{%
Didegah%
\ \BBA {} Thelwall%
}{%
{\protect \APACyear {2013}}%
{\protect \APACexlab {{\protect \BCnt {2}}}}}]{%
didegah2013factors}
\APACinsertmetastar {%
didegah2013factors}%
\begin{APACrefauthors}%
Didegah, F.%
\BCBT {}\ \BBA {} Thelwall, M.%
\end{APACrefauthors}%
\unskip\
\newblock
\APACrefYearMonthDay{2013{\protect \BCnt {2}}}{}{}.
\newblock
{\BBOQ}\APACrefatitle {Which factors help authors produce the highest impact
  research? Collaboration, journal and document properties} {Which factors help
  authors produce the highest impact research? collaboration, journal and
  document properties}.{\BBCQ}
\newblock
\APACjournalVolNumPages{Journal of Informetrics}{7}{4}{861--873}.
\PrintBackRefs{\CurrentBib}

\bibitem [\protect \citeauthoryear {%
Fu%
\ \BBA {} Aliferis%
}{%
Fu%
\ \BBA {} Aliferis%
}{%
{\protect \APACyear {2010}}%
}]{%
fu2010using}
\APACinsertmetastar {%
fu2010using}%
\begin{APACrefauthors}%
Fu, L.%
\BCBT {}\ \BBA {} Aliferis, C.%
\end{APACrefauthors}%
\unskip\
\newblock
\APACrefYearMonthDay{2010}{}{}.
\newblock
{\BBOQ}\APACrefatitle {Using content-based and bibliometric features for
  machine learning models to predict citation counts in the biomedical
  literature} {Using content-based and bibliometric features for machine
  learning models to predict citation counts in the biomedical
  literature}.{\BBCQ}
\newblock
\APACjournalVolNumPages{Scientometrics}{85}{1}{257--270}.
\PrintBackRefs{\CurrentBib}

\bibitem [\protect \citeauthoryear {%
Gl{\"a}nzel%
}{%
Gl{\"a}nzel%
}{%
{\protect \APACyear {1997}}%
}]{%
glanzel1997possibility}
\APACinsertmetastar {%
glanzel1997possibility}%
\begin{APACrefauthors}%
Gl{\"a}nzel, W.%
\end{APACrefauthors}%
\unskip\
\newblock
\APACrefYearMonthDay{1997}{}{}.
\newblock
{\BBOQ}\APACrefatitle {On the possibility and reliability of predictions based
  on stochastic citation processes} {On the possibility and reliability of
  predictions based on stochastic citation processes}.{\BBCQ}
\newblock
\APACjournalVolNumPages{Scientometrics}{40}{3}{481--492}.
\PrintBackRefs{\CurrentBib}

\bibitem [\protect \citeauthoryear {%
Haslam%
\ \protect \BOthers {.}}{%
Haslam%
\ \protect \BOthers {.}}{%
{\protect \APACyear {2008}}%
}]{%
haslam2008makes}
\APACinsertmetastar {%
haslam2008makes}%
\begin{APACrefauthors}%
Haslam, N.%
, Ban, L.%
, Kaufmann, L.%
, Loughnan, S.%
, Peters, K.%
, Whelan, J.%
\BCBL {}\ \BBA {} Wilson, S.%
\end{APACrefauthors}%
\unskip\
\newblock
\APACrefYearMonthDay{2008}{}{}.
\newblock
{\BBOQ}\APACrefatitle {What makes an article influential? Predicting impact in
  social and personality psychology} {What makes an article influential?
  predicting impact in social and personality psychology}.{\BBCQ}
\newblock
\APACjournalVolNumPages{Scientometrics}{76}{1}{169--185}.
\PrintBackRefs{\CurrentBib}

\bibitem [\protect \citeauthoryear {%
Haslam%
\ \BBA {} Koval%
}{%
Haslam%
\ \BBA {} Koval%
}{%
{\protect \APACyear {2010}}%
}]{%
haslam2010predicting}
\APACinsertmetastar {%
haslam2010predicting}%
\begin{APACrefauthors}%
Haslam, N.%
\BCBT {}\ \BBA {} Koval, P.%
\end{APACrefauthors}%
\unskip\
\newblock
\APACrefYearMonthDay{2010}{}{}.
\newblock
{\BBOQ}\APACrefatitle {Predicting long-term citation impact of articles in
  social and personality psychology} {Predicting long-term citation impact of
  articles in social and personality psychology}.{\BBCQ}
\newblock
\APACjournalVolNumPages{Psychological Reports}{106}{3}{891--900}.
\PrintBackRefs{\CurrentBib}

\bibitem [\protect \citeauthoryear {%
Hill%
}{%
Hill%
}{%
{\protect \APACyear {1975}}%
}]{%
hill1975}
\APACinsertmetastar {%
hill1975}%
\begin{APACrefauthors}%
Hill, B\BPBI M.%
\end{APACrefauthors}%
\unskip\
\newblock
\APACrefYearMonthDay{1975}{}{}.
\newblock
{\BBOQ}\APACrefatitle {A simple general approach to inference about the tail of
  a distribution} {A simple general approach to inference about the tail of a
  distribution}.{\BBCQ}
\newblock
\APACjournalVolNumPages{Annals of Statistics}{3}{5}{1163--1174}.
\PrintBackRefs{\CurrentBib}

\bibitem [\protect \citeauthoryear {%
Ke%
}{%
Ke%
}{%
{\protect \APACyear {2013}}%
}]{%
ke2013}
\APACinsertmetastar {%
ke2013}%
\begin{APACrefauthors}%
Ke, W.%
\end{APACrefauthors}%
\unskip\
\newblock
\APACrefYearMonthDay{2013}{}{}.
\newblock
{\BBOQ}\APACrefatitle {A fitness model for scholarly impact analysis} {A
  fitness model for scholarly impact analysis}.{\BBCQ}
\newblock
\APACjournalVolNumPages{Scientometrics}{94}{3}{981--998}.
\PrintBackRefs{\CurrentBib}

\bibitem [\protect \citeauthoryear {%
Koenker%
\ \BBA {} Bassett%
}{%
Koenker%
\ \BBA {} Bassett%
}{%
{\protect \APACyear {1978}}%
}]{%
koenker1978}
\APACinsertmetastar {%
koenker1978}%
\begin{APACrefauthors}%
Koenker, R.%
\BCBT {}\ \BBA {} Bassett, G., Jr.%
\end{APACrefauthors}%
\unskip\
\newblock
\APACrefYearMonthDay{1978}{}{}.
\newblock
{\BBOQ}\APACrefatitle {Regression quantiles} {Regression quantiles}.{\BBCQ}
\newblock
\APACjournalVolNumPages{Econometrica}{46}{1}{33--50}.
\PrintBackRefs{\CurrentBib}

\bibitem [\protect \citeauthoryear {%
Levitt%
\ \BBA {} Thelwall%
}{%
Levitt%
\ \BBA {} Thelwall%
}{%
{\protect \APACyear {2008}}%
}]{%
levitt2008patterns}
\APACinsertmetastar {%
levitt2008patterns}%
\begin{APACrefauthors}%
Levitt, J.%
\BCBT {}\ \BBA {} Thelwall, M.%
\end{APACrefauthors}%
\unskip\
\newblock
\APACrefYearMonthDay{2008}{}{}.
\newblock
{\BBOQ}\APACrefatitle {Patterns of annual citation of highly cited articles and
  the prediction of their citation ranking: A comparison across subjects}
  {Patterns of annual citation of highly cited articles and the prediction of
  their citation ranking: A comparison across subjects}.{\BBCQ}
\newblock
\APACjournalVolNumPages{Scientometrics}{77}{1}{41--60}.
\PrintBackRefs{\CurrentBib}

\bibitem [\protect \citeauthoryear {%
Levitt%
\ \BBA {} Thelwall%
}{%
Levitt%
\ \BBA {} Thelwall%
}{%
{\protect \APACyear {2011}}%
}]{%
levitt2011combined}
\APACinsertmetastar {%
levitt2011combined}%
\begin{APACrefauthors}%
Levitt, J.%
\BCBT {}\ \BBA {} Thelwall, M.%
\end{APACrefauthors}%
\unskip\
\newblock
\APACrefYearMonthDay{2011}{}{}.
\newblock
{\BBOQ}\APACrefatitle {A combined bibliometric indicator to predict article
  impact} {A combined bibliometric indicator to predict article impact}.{\BBCQ}
\newblock
\APACjournalVolNumPages{Information Processing and
  Management}{47}{2}{300--308}.
\PrintBackRefs{\CurrentBib}

\bibitem [\protect \citeauthoryear {%
Lokker%
, McKibbon%
, McKinlay%
, Wilczynski%
\BCBL {}\ \BBA {} Haynes%
}{%
Lokker%
\ \protect \BOthers {.}}{%
{\protect \APACyear {2008}}%
}]{%
lokker2008prediction}
\APACinsertmetastar {%
lokker2008prediction}%
\begin{APACrefauthors}%
Lokker, C.%
, McKibbon, K.%
, McKinlay, R.%
, Wilczynski, N.%
\BCBL {}\ \BBA {} Haynes, R.%
\end{APACrefauthors}%
\unskip\
\newblock
\APACrefYearMonthDay{2008}{}{}.
\newblock
{\BBOQ}\APACrefatitle {Prediction of citation counts for clinical articles at
  two years using data available within three weeks of publication:
  Retrospective cohort study} {Prediction of citation counts for clinical
  articles at two years using data available within three weeks of publication:
  Retrospective cohort study}.{\BBCQ}
\newblock
\APACjournalVolNumPages{BMJ}{336}{7645}{655--657}.
\PrintBackRefs{\CurrentBib}

\bibitem [\protect \citeauthoryear {%
Mingers%
}{%
Mingers%
}{%
{\protect \APACyear {2008}}%
}]{%
mingers2008exploring}
\APACinsertmetastar {%
mingers2008exploring}%
\begin{APACrefauthors}%
Mingers, J.%
\end{APACrefauthors}%
\unskip\
\newblock
\APACrefYearMonthDay{2008}{}{}.
\newblock
{\BBOQ}\APACrefatitle {Exploring the dynamics of journal citations: Modelling
  with S-curves} {Exploring the dynamics of journal citations: Modelling with
  s-curves}.{\BBCQ}
\newblock
\APACjournalVolNumPages{Journal of the Operational Research
  Society}{59}{8}{1013--1025}.
\PrintBackRefs{\CurrentBib}

\bibitem [\protect \citeauthoryear {%
Mingers%
\ \BBA {} Burrell%
}{%
Mingers%
\ \BBA {} Burrell%
}{%
{\protect \APACyear {2006}}%
}]{%
mingers2006modeling}
\APACinsertmetastar {%
mingers2006modeling}%
\begin{APACrefauthors}%
Mingers, J.%
\BCBT {}\ \BBA {} Burrell, Q.%
\end{APACrefauthors}%
\unskip\
\newblock
\APACrefYearMonthDay{2006}{}{}.
\newblock
{\BBOQ}\APACrefatitle {Modeling citation behavior in management science
  journals} {Modeling citation behavior in management science journals}.{\BBCQ}
\newblock
\APACjournalVolNumPages{Information Processing and
  Management}{42}{6}{1451--1464}.
\PrintBackRefs{\CurrentBib}

\bibitem [\protect \citeauthoryear {%
Mingers%
\ \BBA {} Xu%
}{%
Mingers%
\ \BBA {} Xu%
}{%
{\protect \APACyear {2010}}%
}]{%
mingers2010drivers}
\APACinsertmetastar {%
mingers2010drivers}%
\begin{APACrefauthors}%
Mingers, J.%
\BCBT {}\ \BBA {} Xu, F.%
\end{APACrefauthors}%
\unskip\
\newblock
\APACrefYearMonthDay{2010}{}{}.
\newblock
{\BBOQ}\APACrefatitle {The drivers of citations in management science journals}
  {The drivers of citations in management science journals}.{\BBCQ}
\newblock
\APACjournalVolNumPages{European Journal of Operational
  Research}{205}{2}{422--430}.
\PrintBackRefs{\CurrentBib}

\bibitem [\protect \citeauthoryear {%
Onodera%
\ \BBA {} Yoshikane%
}{%
Onodera%
\ \BBA {} Yoshikane%
}{%
{\protect \APACyear {{\protect \BIP {}}}}%
}]{%
onodera_inpress_factors}
\APACinsertmetastar {%
onodera_inpress_factors}%
\begin{APACrefauthors}%
Onodera, N.%
\BCBT {}\ \BBA {} Yoshikane, F.%
\end{APACrefauthors}%
\unskip\
\newblock
\APACrefYearMonthDay{{\protect \BIP {}}}{}{}.
\newblock
{\BBOQ}\APACrefatitle {Factors affecting citation rates of research articles}
  {Factors affecting citation rates of research articles}.{\BBCQ}
\newblock
\APACjournalVolNumPages{Journal of the Association for Information Science and
  Technology}{}{}{}.
\PrintBackRefs{\CurrentBib}

\bibitem [\protect \citeauthoryear {%
Peters%
\ \BBA {} Van~Raan%
}{%
Peters%
\ \BBA {} Van~Raan%
}{%
{\protect \APACyear {1994}}%
}]{%
peters1994determinants}
\APACinsertmetastar {%
peters1994determinants}%
\begin{APACrefauthors}%
Peters, H.%
\BCBT {}\ \BBA {} Van~Raan, A.%
\end{APACrefauthors}%
\unskip\
\newblock
\APACrefYearMonthDay{1994}{}{}.
\newblock
{\BBOQ}\APACrefatitle {On determinants of citation scores: A case study in
  chemical engineering} {On determinants of citation scores: A case study in
  chemical engineering}.{\BBCQ}
\newblock
\APACjournalVolNumPages{Journal of the American Society for Information
  Science}{45}{1}{39--49}.
\PrintBackRefs{\CurrentBib}

\bibitem [\protect \citeauthoryear {%
Redner%
}{%
Redner%
}{%
{\protect \APACyear {1998}}%
}]{%
redner1998}
\APACinsertmetastar {%
redner1998}%
\begin{APACrefauthors}%
Redner, S.%
\end{APACrefauthors}%
\unskip\
\newblock
\APACrefYearMonthDay{1998}{}{}.
\newblock
{\BBOQ}\APACrefatitle {How popular is your paper? An empirical study of the
  citation distribution} {How popular is your paper? an empirical study of the
  citation distribution}.{\BBCQ}
\newblock
\APACjournalVolNumPages{European Physical Journal B}{4}{2}{131--134}.
\PrintBackRefs{\CurrentBib}

\bibitem [\protect \citeauthoryear {%
Stern%
}{%
Stern%
}{%
{\protect \APACyear {2014}}%
}]{%
stern2014high}
\APACinsertmetastar {%
stern2014high}%
\begin{APACrefauthors}%
Stern, D.%
\end{APACrefauthors}%
\unskip\
\newblock
\APACrefYearMonthDay{2014}{}{}.
\newblock
{\BBOQ}\APACrefatitle {High-ranked social science journal articles can be
  identified from early citation information} {High-ranked social science
  journal articles can be identified from early citation information}.{\BBCQ}
\newblock
\APACjournalVolNumPages{PLoS ONE}{9}{11}{e112520}.
\PrintBackRefs{\CurrentBib}

\bibitem [\protect \citeauthoryear {%
Thelwall%
\ \BBA {} Wilson%
}{%
Thelwall%
\ \BBA {} Wilson%
}{%
{\protect \APACyear {{\protect \BIP {}}}}%
}]{%
thelwall_inpress_mendeley}
\APACinsertmetastar {%
thelwall_inpress_mendeley}%
\begin{APACrefauthors}%
Thelwall, M.%
\BCBT {}\ \BBA {} Wilson, P.%
\end{APACrefauthors}%
\unskip\
\newblock
\APACrefYearMonthDay{{\protect \BIP {}}}{}{}.
\newblock
{\BBOQ}\APACrefatitle {Mendeley readership altmetrics for medical articles: An
  analysis of 45 fields} {Mendeley readership altmetrics for medical articles:
  An analysis of 45 fields}.{\BBCQ}
\newblock
\APACjournalVolNumPages{Journal of the Association for Information Science and
  Technology}{}{}{}.
\PrintBackRefs{\CurrentBib}

\bibitem [\protect \citeauthoryear {%
Van~Raan%
}{%
Van~Raan%
}{%
{\protect \APACyear {2004}}%
}]{%
vanraan2004sleeping}
\APACinsertmetastar {%
vanraan2004sleeping}%
\begin{APACrefauthors}%
Van~Raan, A.%
\end{APACrefauthors}%
\unskip\
\newblock
\APACrefYearMonthDay{2004}{}{}.
\newblock
{\BBOQ}\APACrefatitle {Sleeping beauties in science} {Sleeping beauties in
  science}.{\BBCQ}
\newblock
\APACjournalVolNumPages{Scientometrics}{59}{3}{467--472}.
\PrintBackRefs{\CurrentBib}

\bibitem [\protect \citeauthoryear {%
Wallace%
, Larivi{\`e}re%
\BCBL {}\ \BBA {} Gingras%
}{%
Wallace%
\ \protect \BOthers {.}}{%
{\protect \APACyear {2009}}%
}]{%
wallace2009}
\APACinsertmetastar {%
wallace2009}%
\begin{APACrefauthors}%
Wallace, M\BPBI L.%
, Larivi{\`e}re, V.%
\BCBL {}\ \BBA {} Gingras, Y.%
\end{APACrefauthors}%
\unskip\
\newblock
\APACrefYearMonthDay{2009}{}{}.
\newblock
{\BBOQ}\APACrefatitle {Modeling a century of citation distributions} {Modeling
  a century of citation distributions}.{\BBCQ}
\newblock
\APACjournalVolNumPages{Journal of Informetrics}{3}{4}{296--303}.
\PrintBackRefs{\CurrentBib}

\bibitem [\protect \citeauthoryear {%
Walters%
}{%
Walters%
}{%
{\protect \APACyear {2006}}%
}]{%
walters2006predicting}
\APACinsertmetastar {%
walters2006predicting}%
\begin{APACrefauthors}%
Walters, G.%
\end{APACrefauthors}%
\unskip\
\newblock
\APACrefYearMonthDay{2006}{}{}.
\newblock
{\BBOQ}\APACrefatitle {Predicting subsequent citations to articles published in
  twelve crime-psychology journals: Author impact versus journal impact}
  {Predicting subsequent citations to articles published in twelve
  crime-psychology journals: Author impact versus journal impact}.{\BBCQ}
\newblock
\APACjournalVolNumPages{Scientometrics}{69}{3}{499--510}.
\PrintBackRefs{\CurrentBib}

\bibitem [\protect \citeauthoryear {%
D.~Wang%
, Song%
\BCBL {}\ \BBA {} Barab{\'a}si%
}{%
D.~Wang%
\ \protect \BOthers {.}}{%
{\protect \APACyear {2013}}%
}]{%
wang2013quantifying}
\APACinsertmetastar {%
wang2013quantifying}%
\begin{APACrefauthors}%
Wang, D.%
, Song, C.%
\BCBL {}\ \BBA {} Barab{\'a}si, A\BHBI L.%
\end{APACrefauthors}%
\unskip\
\newblock
\APACrefYearMonthDay{2013}{}{}.
\newblock
{\BBOQ}\APACrefatitle {Quantifying long-term scientific impact} {Quantifying
  long-term scientific impact}.{\BBCQ}
\newblock
\APACjournalVolNumPages{Science}{342}{6154}{127--132}.
\PrintBackRefs{\CurrentBib}

\bibitem [\protect \citeauthoryear {%
J.~Wang%
}{%
J.~Wang%
}{%
{\protect \APACyear {2013}}%
}]{%
wang2013citation}
\APACinsertmetastar {%
wang2013citation}%
\begin{APACrefauthors}%
Wang, J.%
\end{APACrefauthors}%
\unskip\
\newblock
\APACrefYearMonthDay{2013}{}{}.
\newblock
{\BBOQ}\APACrefatitle {Citation time window choice for research impact
  evaluation} {Citation time window choice for research impact
  evaluation}.{\BBCQ}
\newblock
\APACjournalVolNumPages{Scientometrics}{94}{3}{851--872}.
\PrintBackRefs{\CurrentBib}

\bibitem [\protect \citeauthoryear {%
M.~Wang%
\ \protect \BOthers {.}}{%
M.~Wang%
\ \protect \BOthers {.}}{%
{\protect \APACyear {2012}}%
}]{%
wang2012development}
\APACinsertmetastar {%
wang2012development}%
\begin{APACrefauthors}%
Wang, M.%
, Yu, G.%
, Xu, J.%
, He, H.%
, Yu, D.%
\BCBL {}\ \BBA {} An, S.%
\end{APACrefauthors}%
\unskip\
\newblock
\APACrefYearMonthDay{2012}{}{}.
\newblock
{\BBOQ}\APACrefatitle {Development a case-based classifier for predicting
  highly cited papers} {Development a case-based classifier for predicting
  highly cited papers}.{\BBCQ}
\newblock
\APACjournalVolNumPages{Journal of Informetrics}{6}{4}{586--599}.
\PrintBackRefs{\CurrentBib}

\bibitem [\protect \citeauthoryear {%
M.~Wang%
, Yu%
\BCBL {}\ \BBA {} Yu%
}{%
M.~Wang%
\ \protect \BOthers {.}}{%
{\protect \APACyear {2011}}%
}]{%
wang2011mining}
\APACinsertmetastar {%
wang2011mining}%
\begin{APACrefauthors}%
Wang, M.%
, Yu, G.%
\BCBL {}\ \BBA {} Yu, D.%
\end{APACrefauthors}%
\unskip\
\newblock
\APACrefYearMonthDay{2011}{}{}.
\newblock
{\BBOQ}\APACrefatitle {Mining typical features for highly cited papers} {Mining
  typical features for highly cited papers}.{\BBCQ}
\newblock
\APACjournalVolNumPages{Scientometrics}{87}{3}{695--706}.
\PrintBackRefs{\CurrentBib}

\bibitem [\protect \citeauthoryear {%
Yu%
, Yu%
, Li%
\BCBL {}\ \BBA {} Wang%
}{%
Yu%
\ \protect \BOthers {.}}{%
{\protect \APACyear {2014}}%
}]{%
yu2014citation}
\APACinsertmetastar {%
yu2014citation}%
\begin{APACrefauthors}%
Yu, T.%
, Yu, G.%
, Li, P\BHBI Y.%
\BCBL {}\ \BBA {} Wang, L.%
\end{APACrefauthors}%
\unskip\
\newblock
\APACrefYearMonthDay{2014}{}{}.
\newblock
{\BBOQ}\APACrefatitle {Citation impact prediction for scientific papers using
  stepwise regression analysis} {Citation impact prediction for scientific
  papers using stepwise regression analysis}.{\BBCQ}
\newblock
\APACjournalVolNumPages{Scientometrics}{101}{2}{1233--1252}.
\PrintBackRefs{\CurrentBib}

\end{thebibliography}
\end{document}